\documentclass[a4paper,10pt]{article}
\usepackage[utf8x]{inputenc}
\usepackage{amssymb}
\usepackage{amsmath}
\usepackage{graphicx}     
\usepackage[a4paper]{geometry}
\usepackage{upgreek}
\usepackage{subfig}
\usepackage{fixmath}
\usepackage{lmodern}
\usepackage{caption}
\usepackage{color}
\usepackage{hyperref}
\usepackage[sort&compress]{natbib}
\bibpunct{}{}{,}{s}{,}{,}
\usepackage{todonotes}

\makeatletter
 \renewcommand{\p@subfigure}{}
\makeatother
\makeatletter
 \providecommand\phantomcaption{\caption@refstepcounter\@captype}
\makeatother
\DeclareCaptionFormat{PoP}{hcode using #1, #2 and #3}

\renewcommand{\epsilon}{\varepsilon}

\newcommand{\grad}{\nabla}

\newcommand{\D}[2]{\ensuremath{\frac{\partial#2}{\partial#1}}}
\renewcommand{\vec}[1]{\ensuremath{\boldsymbol{#1}}} % bold vectors, instead
\newcommand{\figplacing}{hbt}
\newcommand{\figwidth}{0.80\linewidth}
\newcommand{\iu}{\ensuremath{\mathrm{i}}}
\newcommand{\note}[1]{\ensuremath{\mathrm{^#1}}}

\usepackage{sectsty}
\allsectionsfont{\sffamily}

\title{Impurity transport in temperature gradient driven turbulence}
\author{Andreas Skyman\footnote{Euratom--VR Association, Department of Earth and Space
      Sciences, Chalmers University of Technology, SE-412 96 Göteborg, Sweden},
      H. Nordman\footnotemark[1], P. Strand\footnotemark[1]}

\begin{document}

%% Remove this in final version:
%{\noindent \Large \sffamily \bfseries Impurity transport in temperature gradient driven turbulence\\ \today}
%
%
%\listoftodos
%\TODO{Remove TODO-list in final version}
%\TODO{a final spell check}
%\tableofcontents
%\thispagestyle{empty}
%\newpage
%~
%\thispagestyle{empty}
%\cleardoublepage
%\setcounter{page}{1}

%\maketitle
{\noindent \Large \sffamily \bfseries Impurity transport in temperature gradient driven turbulence}

~

{\large \sffamily A. Skyman,\footnotemark[1] H. Nordman,\footnotemark[1] and P. Strand\footnotemark[1]}

{\textit{\small{\footnotemark[1]Euratom--VR Association, Department of Earth and Space Sciences, \\\indent Chalmers University of Technology, SE-412 96 Göteborg, Sweden}}}

\renewcommand{\thefootnote}{\alph{footnote})} 
\begin{abstract}
\noindent 
In the present paper the transport of impurities driven by trapped electron~(TE) mode turbulence is studied.
Non-linear~(NL) gyrokinetic simulations using the code GENE are compared with results from quasilinear~(QL) gyrokinetic simulations and a computationally efficient fluid model.
The main focus is on model comparisons for electron temperature gradient driven turbulence regarding the sign of the convective impurity velocity~(pinch) and the impurity density gradient $R/L_{n_Z}$ (peaking factor) for zero impurity flux.
In particular, the scaling of the impurity peaking factors with impurity charge $Z$ and with driving temperature gradient is investigated and compared with the results for Ion Temperature Gradient (ITG) driven turbulence.
In addition, the impurity peaking is compared to the main ion peaking obtained by a self-consistent fluid calculation of the density gradients corresponding to zero particle fluxes. 

For the scaling of the peaking factor with impurity charge $Z$, a weak dependence is obtained from NL~GENE and fluid simulations.
The QL GENE results show a stronger dependence for low $Z$ impurities and overestimates the peaking factor by up to a factor of two in this region.
As in the case of ITG dominated turbulence, the peaking factors saturate as $Z$ increases, at a level much below neoclassical predictions.
However, the scaling with $Z$ is weak or reversed as compared to the ITG case.

The scaling of impurity peaking with the background temperature gradients is found to be weak in the NL~GENE and fluid simulations.
The QL results are also here found to significantly overestimate the peaking factor for low $Z$ values.

For the parameters considered, the background density gradient for zero particle flux is found to be slightly larger than the corresponding impurity zero flux gradient.
\end{abstract}

\section{Introduction} \label{sec:intro}
The transport properties of impurities is of high relevance for the performance and optimisation of magnetic fusion devices. 
For instance, the possible accumulation of He ash in the core of the reactor plasma will serve to dilute the fuel, thus lowering fusion power.
Heavier impurity species, originating from the plasma-facing surfaces, may also accumulate in the core, and wall-impurities of relatively low density may lead to unacceptable energy losses in the form of radiation.\cite{Harte2010}
In an operational power plant, both impurities of low and high charge numbers will be present.
% and both  may be detrimental to the fusion process.

In the confinement zone of tokamaks, the transport of the background species is usually dominated by turbulence.
The Trapped Electron~(TE) mode and the Ion Temperature Gradient~(ITG\footnote{also commonly referred to as the $\eta_i$ mode}) mode are expected to be the main contributors. 
Turbulent impurity transport has been investigated in a number of theoretical\cite{Frojdh1992, Basu2003, Estrada-Mila2005, Naulin2005, Priego2005, Fulop2006, Bourdelle2007, Dubuit2007, Camenen2009, Fulop2010, Futatani2010, Hein2010, Moradi2010, Fulop2011, Nordman2008, Angioni2006, Angioni2007, Nordman2007a, Angioni2009a, Fulop2009, Nordman2011} and experimental\cite{Dux2003, Puiatti2003,  Puiatti2006, Giroud2007, Giroud2009} papers.
In tokamak experiments, also the impurity transport is usually dominated by turbulence, resulting in impurity peaking factors well below the neoclassical predictions.\cite{Helander2002, Angioni2006, Angioni2007, Nordman2007a, Angioni2009a, Fulop2009, Giroud2009, Nordman2011}
The main theoretical effort has, with a few exceptions,\cite{Angioni2009a, Angioni2011, Dubuit2007, Estrada-Mila2005} hitherto been devoted to quasilinear studies, primarily focused on ITG mode driven impurity transport.
For the directly reactor relevant regimes, however, where $\alpha$-particle heating dominates, as will be the case in the ITER device, or in electron cyclotron resonance heated plasmas, TE~mode driven impurity transport will likely be important. 

In the present study, transport of impurities driven by TE mode turbulence is investigated by NL gyrokinetic simulations using the code GENE.\cite{Jenko2000, Dannert2005a, Merz2008a}
The simulation results are compared to QL gyrokinetic simulations as well as results obtained from a multi-fluid model.\cite{Weiland2000}
The fluid model is employed for the dual purposes of benchmarking a computationally efficient model, suitable for predictive simulations, and interpreting the results.
The TE~mode results are compared with the more well known results for ITG~mode dominated turbulence, obtained from QL gyrokinetic and fluid simulations.

The impurity diffusivity ($D_Z$) and convective velocity ($V_Z$) are estimated from simulation data, and from these the zero-flux peaking factor impurity density gradient $\left(R/L_{n_Z}=-RV_Z/D_Z\right)$, also referred to as the impurity peaking factor ($PF$) is derived. 
This quantity expresses the impurity density gradient at which the convective and diffusive transport of impurities are exactly balanced.
The sign of $PF$ is of special interest, as it determines whether the impurities are subject to an inward ($PF > 0$) or outward ($PF < 0$) pinch. 
Scalings of peaking factors with impurity charge ($Z$), electron and ion temperature gradients ($\grad T_{e,i}$), and electron density gradient ($\grad n_e$), are studied, giving particular attention to $\grad T_e$ driven TE~mode impurity transport.
The results are compared and contrasted with results from previous studies focused on ITG driven impurity transport.
In addition, the impurity peaking relative to the main ion peaking in the plasma core obtained from a self-consistent treatment of the particle fluxes will be discussed.
This section is experimentally relevant in situations with edge particle fuelling where the steady state gradient corresponds to zero particle flux.

The remainder of the paper is structured as follows: first the transport models are reviewed, beginning with of the fluid model employed (section~\ref{sec:fluid}) where the focus is on the impurity dynamics. 
This is followed by a brief introduction of the gyrokinetic model and the GENE code (section~\ref{sec:gyro}) and a section on the specifics of the simulations (section~\ref{sec:simulations}). 
After this, the main results are covered, including discussion and interpretation of the acquired results (section~\ref{sec:results}).
The final section of the paper is a summary of the main conclusions to be drawn (section~\ref{sec:conclusions}).

\section{Transport models}
%To arrive at a set of equations describing the impurity transport that are both meaningful and solvable, some approximations are necessary. 
%The advances in high performance computing have allowed fusion modellers to move from fluid descriptions of the plasma to kinetic descriptions as the basis for turbulence modelling, however, the underlying physics are easier to grasp from fluid models.

%In kinetic theory the plasma is described through distribution functions of velocity and position for each of the included plasma species.  
%Hence, kinetic equations are inherently six-dimensional, however, in magnetically confined fusion plasmas the confined particles are generally constrained to tight orbits along field lines.  
%This motivates averaging over the gyration, reducing the problem to five-dimensional gyrokinetic equations.\cite{Antonsen1980,Frieman1982,Hahm1988,Brizard1989}
%Since the equations governing the evolution of the distributions are all coupled, the resulting decrease in numerical complexity is considerable.

%Fluid theory, on the other hand, is derived by taking the moments of the kinetic equations to some order, making them tractable by finding an appropriate closure.\cite{Weiland2000}
%In addition to making the workings of the plasma more accessible, by reintroducing familiar physical concepts such as pressure and density, fluid models are also several orders of magnitude more computationally efficient. 

\subsection{Fluid theory} \label{sec:fluid}
The Weiland multi-fluid model\cite{Weiland2000} consists of coupled sets of equations for each constituent particle species: main ions, electrons and impurities.\cite{Frojdh1992, Nordman2007a, Nordman2007b, Nordman2008, Nordman2011} 
Effects of toroidal rotation is not included here.
Neglecting finite Larmor radius effects, the impurity equations of continuity, parallel motion and energy take the form of Eq.~\eqref{eq:continuity}--\eqref{eq:energy}.

\begin{eqnarray} %eq:continuity, eq:parallelmotion, eq:energy
\label{eq:continuity} 
 \left(\widetilde{\omega} + \tau_Z^*\right)\widetilde{n}_Z -
\left(\frac{R}{2 L_{n_Z}} - \lambda\right)\widetilde{\phi} + 
 \tau^*_Z\widetilde{T}_Z - \frac{k_\parallel \delta v_{\parallel Z}}{\omega_{D_e}} = 0\\
 \label{eq:parallelmotion}
 \left(\widetilde{\omega} - 2 \tau^*_Z\right)\frac{k_\parallel \delta
v_{\parallel Z}}{\omega_{D_e}} = \frac{Z}{A_Zq_*^2}\widetilde{\phi} +
 \frac{\tau_Z}{A_Zq_*^2}\left(\widetilde{n}_Z + \widetilde{T}_Z\right) \\
 \label{eq:energy}
 \left(\widetilde{\omega} + \frac{5}{3}\tau_Z^*\right)\widetilde{T}_Z - 
 \left(\frac{R}{2L_{T_Z}} -
\frac{1}{3}\frac{R}{L_{n_Z}}\right)\widetilde{\phi} -
\frac{2}{3}\widetilde{\omega}\widetilde{n}_Z = 0
\end{eqnarray}

\noindent In Eq.~\eqref{eq:continuity}--\eqref{eq:energy}, $\widetilde{n}_Z=\delta n_z/n_Z$ is the density,  $\widetilde{\phi}=e\phi/T_e$ the electrostatic potential, $\widetilde{T}_Z=\delta T_Z/T_Z$ the temperature and $\delta v_{\parallel Z}$ the parallel velocity. 
The normalised eigenvalue and wave vector of the eigenmodes are  $\widetilde{\omega}=\widetilde{\omega}_r + \iu\gamma$ and $\vec{k}=k_\parallel \hat{\vec{z}} + \vec{k}_\perp$, ~$\widetilde{ }$ denoting normalisation with respect to the electron magnetic drift frequency $\omega_{De}$.
The normalised scale lengths can be assumed to be constant for the flux tube domain considered, and are defined as $\frac{R}{L_{X_j}}=-\frac{R}{X_j}\D{r}{X_j}$, where $R$ is the major radius of the tokamak, $X_j=n_j,\,T_j$ for species $j$.
The other parameters are defined as follows: $\tau_Z^*=\lambda T_Z/Z T_e$ with $\lambda=\cos\theta + s\theta\sin\theta$ for the poloidal angle $\theta$, $\tau_Z=T_Z/T_e$, $A_Z=m_Z/m_i\approx2 Z$ is the impurity mass number, $Z$ is the impurity charge. 
Further, $s$ is the magnetic shear and $q^*=2 q k_\theta\rho_s$, where $q$ is the safety factor, $\rho_s=c_s/\Omega_{ci}$ is the ion sound scale with the ion sound speed $c_s=\sqrt{T_e/m_i}$ and the ion cyclotron frequency $\Omega_{ci}=eB/m_i$. 
Effects of curvature enter the equations through the magnetic drift, defined as $\omega_{D_Z}=\omega_{D_Z}^{\theta=0}\lambda(\theta)$, which originates from the compression of the $\vec{E}\times\vec{B}$ drift velocity, the diamagnetic drift velocity and the diamagnetic heat flow. 
Curvature effects from the stress tensor enter as $2\tau_Z^*$ at the left hand side of Eq.~\eqref{eq:parallelmotion}.

Combining Eq.~\eqref{eq:continuity}--\eqref{eq:energy}, while neglecting pressure perturbations in Eq.~\eqref{eq:parallelmotion} for simplicity, the relation of the electrostatic potential $\widetilde{\phi}$ and impurity density $\widetilde{n}_Z$ becomes: 
\begin{equation} \label{eq:n_phi}
 \widetilde{n}_Z = \left[\widetilde{\omega}\left(\frac{R}{2 L_{n_Z}}
-\lambda\right) - 
 \tau_Z^*\left(\frac{R}{2 L_{T_Z}} - \frac{7}{3}\frac{R}{2 L_{n_Z}} +
\frac{5 \lambda}{3}\right) + 
 \frac{Z}{A_Z q_*^2}\left(\frac{\widetilde{\omega} +
5 \tau_Z^*/3}{\widetilde{\omega}- 2 \tau_Z^*}\right)
 \right]\frac{\widetilde{\phi}}{N}, 
\end{equation}

\noindent where

\begin{equation} \label{eq:N}
 N=\widetilde{\omega}^2 + \frac{10 \tau_Z^*}{3}\widetilde{\omega} + \frac{5 \tau_Z^{*2}}{3}.
\end{equation}

The main ion and electron response is calculated from the corresponding fluid equations for ions and trapped electrons.
The electron response is given by a trapped and a free part such that $\frac{\delta n_{e}}{n_e} = f_{t}\frac{\delta n_{e_{t}}}{n_{e_{t}}} + \left(1 - f_{t}\right)\frac{e \phi}{T_e}$, i.e. the free electrons are assumed to be adiabatic and thus to follow the Boltzmann distribution: $\delta n_{e_{f}}/n_{e_{f}}=e \phi/T_e$.

The equations are closed by the assumption of quasi-neutrality:
\begin{equation} \label{eq:qneutrality}
 \frac{\delta n_e}{n_e} = \left(1 - Z f_Z\right) \frac{\delta n_i}{n_i} + Z
 f_Z \frac{\delta n_Z}{n_Z},
\end{equation}
\noindent where $f_Z = \frac{n_Z}{n_e}$ is the fraction of impurities.

Thus an eigenvalue equation for TE~and ITG~modes is obtained in the presence of impurities.
Assuming a strongly ballooning eigenfunction with\cite{Hirose1994} $k_\parallel^2=\left(3 q^2 R^2\right)^{-1}$, the eigenvalue equation is reduced to a system of algebraic equations that is solved numerically.
The sensitivity of the fluid results ti the choice of $k_\parallel$ will be examined in section~\ref{sec:kpar} below.

\begin{figure}[\figplacing] % fig:Gamma 
 \centering
 \includegraphics[width=\figwidth]{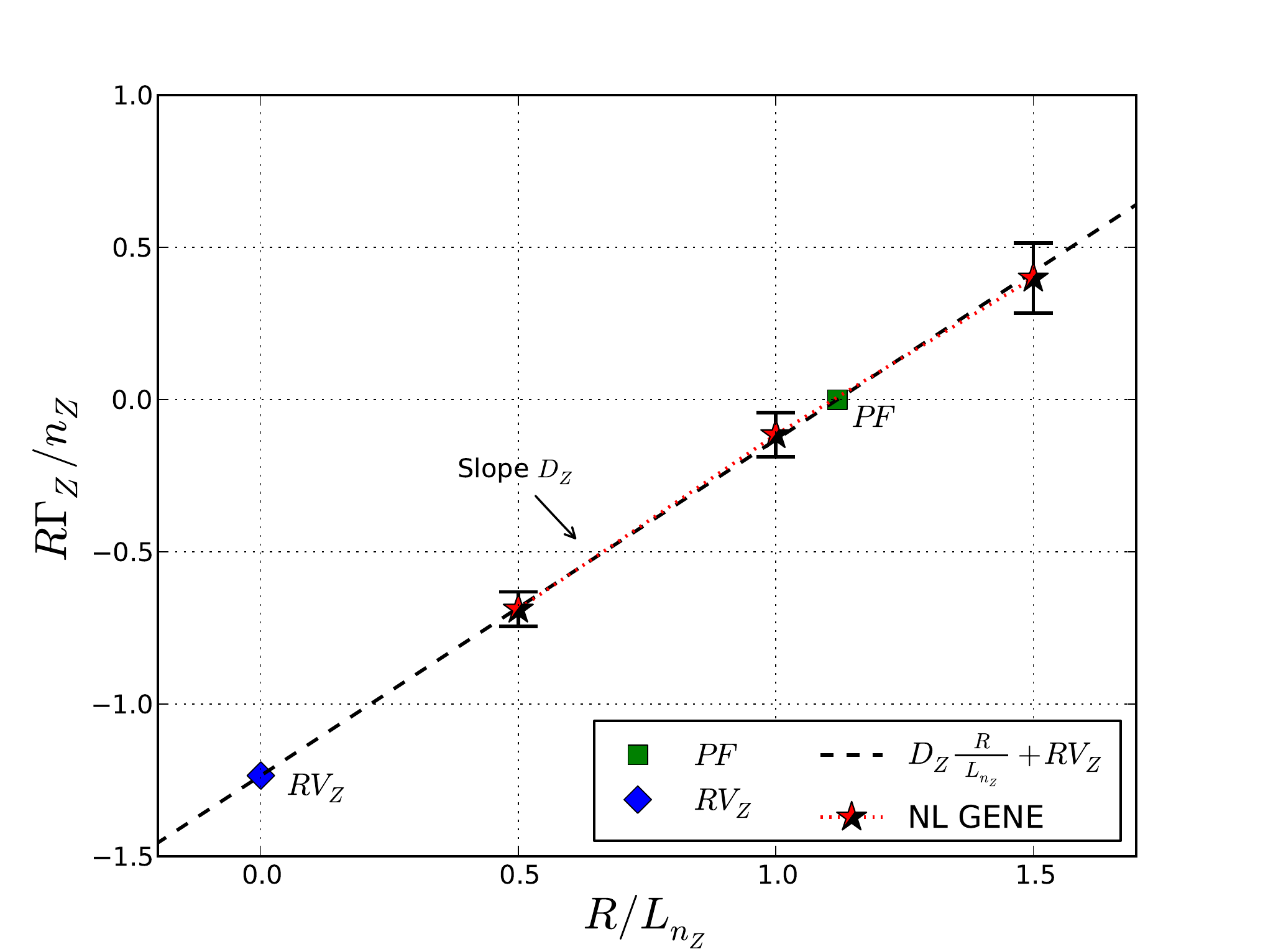}
 \caption{Impurity flux $(\Gamma_Z)$ dependence on the impurity density gradient ($-R\grad n_Z/n_Z=R/L_{n_Z}$), illustrating the peaking factor ($PF$), the diffusivity ($D_Z$) and pinch ($RV_Z$), and the validity of the linearity assumption in Eq.~\eqref{eq:transport_short} of $\Gamma_Z$ for trace impurities. 
 Parameters of Eq.~\eqref{eq:transport_short} are estimated from the calculated fluxes, taking the estimated error of the data into account.
 The flux is acquired as the average of a time series after convergence, as is illustrated in Fig.~\ref{fig:timetrace}.
 Data from NL~GENE simulations of TE~mode driven turbulence with He impurities and parameters as in Fig.~\ref{fig:Z_TEM}. 
 The error bars indicate an estimated error of one standard deviation.}
 \label{fig:Gamma}
\end{figure}

%\subsubsection{Impurity transport} 
\label{sec:PF}
The zero-flux impurity peaking factor, defined as $PF=-\frac{R V_Z}{D_Z}$ for the value of the impurity density gradient that give zero impurity flux, quantifies the balance of convective and diffusive impurity transport. 
Its derivation relies on the fact that the transport of a trace impurity species can be described locally by a diffusive and a convective part.
In the trace impurity limit, i.e. for $Z f_Z \rightarrow 0$ in Eq.~\ref{eq:qneutrality}, the impurity flux $\Gamma_Z$ becomes a linear function of $\grad n_Z$, offset by a convective velocity or ``pinch'' $V_Z$.
The resulting expression can be seen in Eq.~\eqref{eq:transport_short}, where $n_Z$ is the density of the impurity species and $R$ is the major radius of the tokamak, and both the diffusion coefficient ($D_Z$) and the convective velocity ($V_Z$) are independent of $\grad n_Z$.\cite{Angioni2006}
Setting $\Gamma_Z = 0$ in Eq.~\eqref{eq:transport_short} yields the interpretation of $PF$ as the gradient of zero-impurity flux.

\begin{equation} \label{eq:transport_short}
\Gamma_{n_Z} = 
-D_Z\grad n_Z + n_Z V_Z \Leftrightarrow \frac{R\Gamma}{n_Z} =
D_Z\frac{R}{L_{n_Z}} + RV_Z,
\end{equation}

\noindent where the right hand side of the equivalence is arrived at by assuming $-\grad n_Z = 1/L{n_Z}$. The relationship of $PF$ to $D_Z$ and $V_Z$ is illustrated in Fig.~\ref{fig:Gamma}. 

The impurity particle flux at the left hand side of Eq.~\eqref{eq:transport_short} can be written as:

\begin{equation} \label{eq:Gamma_derivation}
\Gamma_{n_Z} =
\left<\delta n_Z v_{\vec{E}\times\vec{B}}\right> = 
-n_Z \rho_s c_s\left<\widetilde{n}_Z\frac{1}{r}\D{\theta}{\widetilde{\phi}}
\right>.
\end{equation}

\noindent The angled brackets imply a time and space average over all unstable modes. Performing this averaging for a fixed length scale $k_\theta\rho_s$ of the turbulence, the following expression is reached:

\begin{multline} \label{eq:transport}
 \frac{\Gamma_{n_Z}}{n_Z  c_s} =
\frac{k_\theta \rho_s \widetilde{\gamma}\left|\widetilde{\phi}_k\right|^2}{
\left|N\right|^2}
 \Biggl\{\left.
 \frac{R}{2 L_{n_Z}}\left(\left|\widetilde{\omega}\right|^2 +
\frac{14 \tau_Z^*}{3} \widetilde{\omega}_r + \frac{55 \tau_Z^{*2}}{9}\right)-
 \right.\\
 \frac{R}{2 L_{T_Z}}\left(2 \tau_Z^* \widetilde{\omega}_r +
\frac{10 \tau_Z^{*2}}{3}\right)-
 \left<\lambda\right>\left(\left|\widetilde{\omega}\right|^2 +
\frac{10 \tau_Z^*}{3} \widetilde{\omega}_r + \frac{35 \tau_Z^{*2}}{9}\right)+
 \\ \left. 
 \frac{Z}{3 A_Z q_*^2 \left|N_1\right|^2}
 \left[ \tau_Z^*\left(
 \frac{19}{3} \widetilde{\omega}_r^2 - \frac{1}{3} \widetilde{\gamma}^2 +
\frac{100 \tau_Z^*}{3} \widetilde{\omega}_r -5 \tau_Z^{*2}
 \right) + 2 \widetilde{\omega}_r\left|\widetilde{\omega}\right|^2 \right]
 \right.\Biggr\},
\end{multline}

\noindent where $N_1=\widetilde{\omega}-2 \tau_Z^*$ is introduced.

In the following it is assumed that the turbulence is isotropic in the radial and poloidal directions ($r$ and $\theta$ respectively; $k_r\rho_s=k_\theta\rho_s$), with a saturated fluctuation level of $\left|\phi_k\right|=\frac{\gamma}{\omega_{*e}}\frac{1}{k_\theta L_{n_e}}$.\cite{Weiland2000}
A brief review of the different mechanisms responsible for the impurity transport, as identified in previous studies\cite{Angioni2006, Estrada-Mila2005, Dubuit2007} is given here.
The first term in Eq.~\eqref{eq:transport} corresponds to the diffusive part of Eq.~\eqref{eq:transport_short}, whereas the three subsequent terms correspond to the convective part of the transport of the impurity species. 
Of this, the $R/L_{T_Z}$ term is the thermodiffusion, the sign of which is governed mainly by the real frequency, $\widetilde{\omega}_r$. 
For TE~modes, $\widetilde{\omega}_r>0$, and for ITG~modes $\widetilde{\omega}_r<0$, resulting the thermodiffusion generally giving an inward pinch for TE~modes and an outward pinch for ITG~modes.
Due to the $Z$-dependence in $\tau_Z^*$, this term scales as $V_Z^{\grad T}\sim (1/Z)(R/L_{T_Z})$ to leading order, rendering it unimportant for large $Z$ impurity species, but it is important for lighter elements, such as the Helium ash.
Further, the $\left<\lambda\right>$ term gives the curvature pinch, which is usually inward, and the final term is the parallel compression term for the impurities.
As opposed to the thermodiffusion, the parallel compression pinch is usually outward for TE~modes and inward for ITG~modes.
Its $Z$ dependence is $V_Z^\parallel\sim Z/A_Z k_\parallel^2\sim Z/A_Z q^2$, but since $A_Z\approx2 Z$ this is is expected to be a very weak scaling.\cite{Angioni2006}
Effects of toroidal rotation on the impurity transport has recently been studied,\cite{Angioni2011, Camenen2009} but will not be considered here.

\subsection{Gyrokinetics -- the GENE code} \label{sec:gyro}
The GENE code\cite{GENE, Jenko2000, Dannert2005a, Merz2008a} is a massively parallel gyrokinetic Vlasov code, solving the nonlinear time evolution of the gyrokinetic distribution functions on a fixed grid in phase space. 
The gyrokinetic equations are derived from the kinetic equations by performing an average of the particles' gyrations around the field lines, so that the equations follow the centre of gyration, rather than the explicit orbits.\cite{Antonsen1980,Frieman1982,Hahm1988,Brizard1989}
This reduces the velocity space coordinates from three to two directions: parallel and perpendicular velocity.
Following the conventions of GENE, these are referred to as $v$ and $\mu$ respectively.
In real space, the radial ($x$) and bi-normal ($y$) dependencies are treated spectrally, i.e. those directions are discretised explicitly in $k$-space, whereas the toroidal ($z$) direction is discretised in real space.
Because all phase space coordinates are coupled nonlinearly, the decrease from six to five phase space coordinates means a significant increase in computational efficiency.
This simplification of the equations is appropriate if the gyro-radii are small compared to the turbulent features, and the cyclotron frequencies large compared to the frequencies of the turbulent phenomena, conditions which generally hold in core tokamak plasmas.\cite{Antonsen1980,Frieman1982,Hahm1988,Brizard1989}

\begin{figure}[\figplacing] %fig:phi-torus
  \centering
  \includegraphics[width=\figwidth]{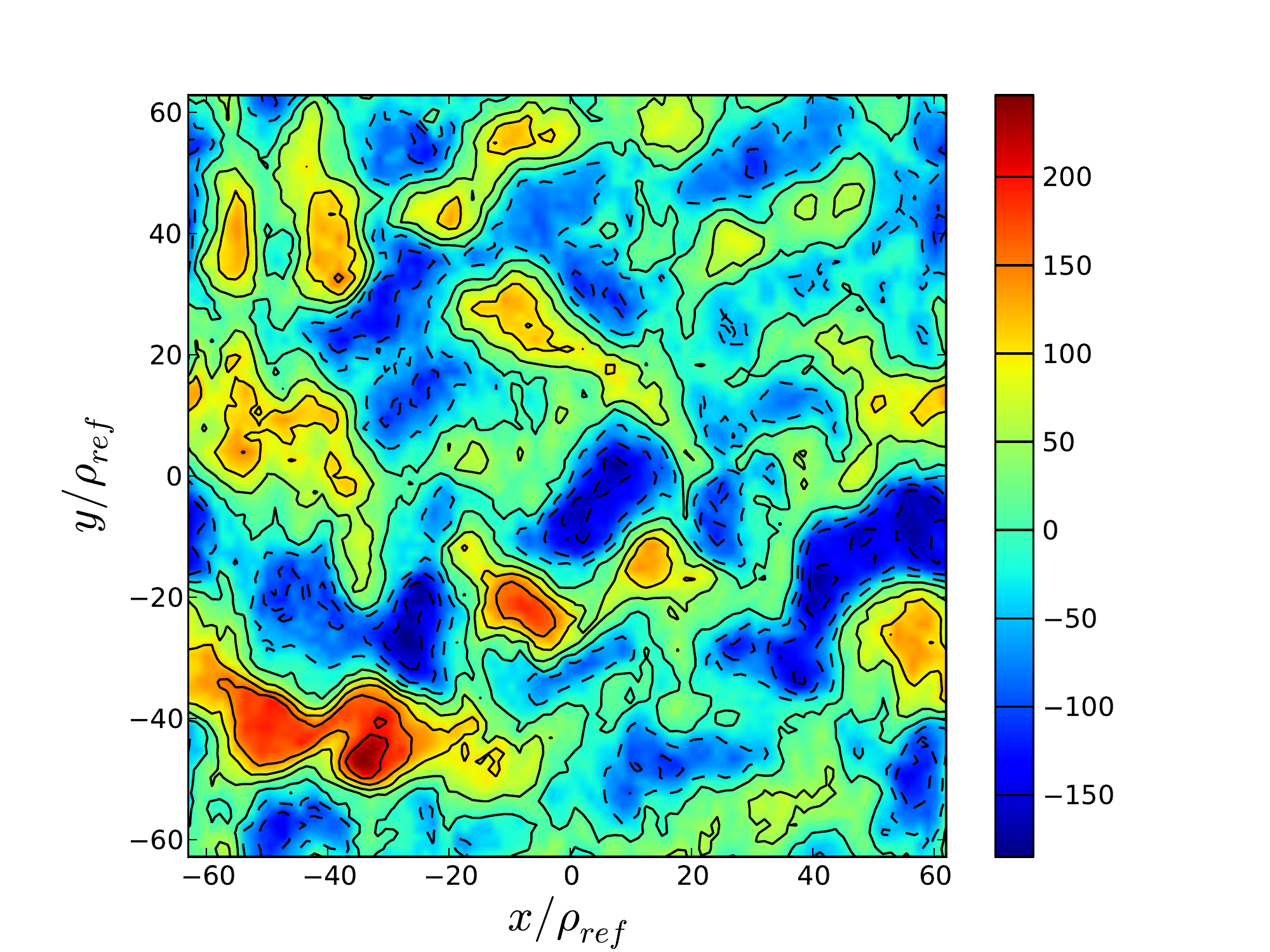}
  \caption{A cross-section of the flux tube, showing the fluctuation of the electrostatic potential $\phi$. Data from NL~GENE simulation of TE~mode turbulence, with parameters as in Fig.~\ref{fig:Z_TEM} at $t\approx300\,R/c_s$.}
  \label{fig:phi-torus}
 \end{figure}

In this paper, GENE simulations are performed in a flux tube geometry with periodic boundary conditions in the perpendicular directions.
The flux tube is in essence a box that is elongated and twisted along with the $\vec{B}$ field as the field lines traverse the tokamak.
Its application relies on the assumption that the scales of the phenomena of interest are all small compared to the size of the whole plasma.
This is generally held to be true in the core of the plasma. 
A cross-section of the flux tube is presented in Fig.~\ref{fig:phi-torus}. 
There the size of the turbulent features can be seen, and a comparison of their size to the flux tube's perpendicular resolution of $\sim125\times125$ main ion gyro-radii and the overall box size indicates that the resolution and flux tube dimensions are adequate; see Section~\ref{sec:simulations} for more details on how the resolution was chosen.

\begin{figure}[\figplacing] % fig:timetrace
 \centering
 \includegraphics[width=\figwidth]{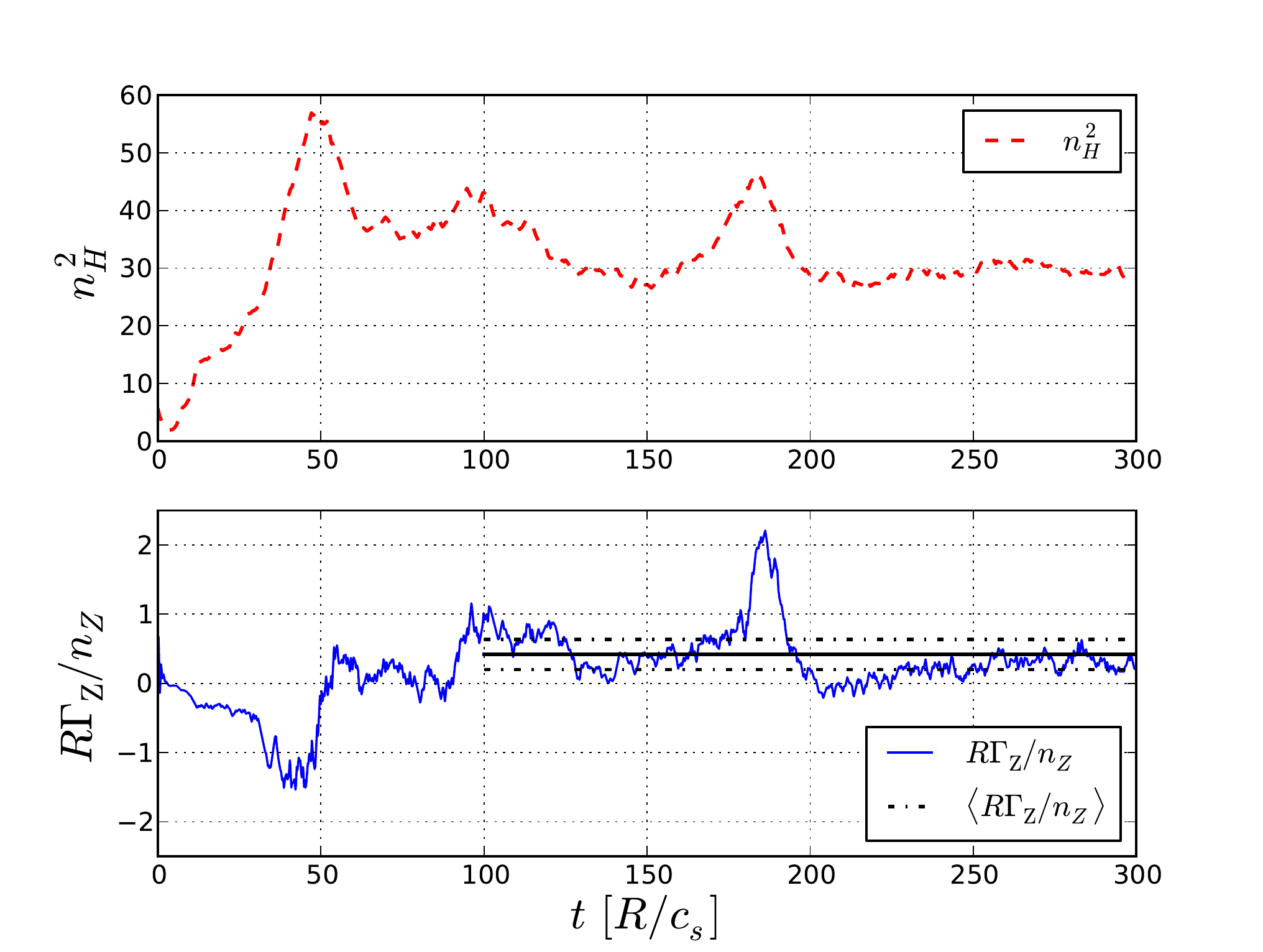}
 \caption{Time series showing fluctuations in the main ion density ($n_H^2$) and impurity flux ($\Gamma_z$) after averaging over the whole flux tube (see Fig.~\ref{fig:phi-torus}).
 The averaged impurity flux ($\left<\Gamma_Z\right>$) is calculated from $\Gamma_Z$, discarding the first portion to ensure that the linear phase of the simulation is not included.
 $\left<\Gamma_Z\right>$ is used for finding the peaking factor for the impurity species, as is illustrated in Fig.~\ref{fig:Gamma}.
 The bursty nature of the transport is seen in the peak around $t\approx 185\,R/c_s$.
 These bursts have been found to affect the average flux little, but to significantly increase the estimated error in $\left<\Gamma_Z\right>$ ($-\cdot-$).
 Data from NL~GENE simulation of TE~mode driven turbulence with He~impurities.
 The parameters are the same as in Fig.~\ref{fig:Z_TEM}, with $-R\grad n_Z/n_Z=R/L_{n_Z}= 1.5$.}
 \label{fig:timetrace}
\end{figure}

The data presented in Fig.~\ref{fig:phi-torus} is computed from the raw field data. 
By integrating further, scalar quantities can be obtained; those quantities are often the most interesting from a physics perspective, since they are easier to compare both to theoretical, experimental, and other numerical results.
In this study, the scalar impurity flux $\Gamma_Z$ is of most interest. 
%, since it is used for deriving the zero-flux peaking factor $PF$ (see section~\ref{sec:PF}). 
Time series showing the fluctuations in the main ion density and the impurity flux for a nonlinear GENE simulation are presented in Fig.~\ref{fig:timetrace}.

GENE can also be run in quasilinear mode, a method that is considerably less demanding when it comes computer resources, since the non-linear coupling between length scales is ignored.\cite{Dannert2005a, Dannert2005b, Merz2008a}
The method used here only captures the contribution from the most unstable, not sub-dominant modes, and only for the particular length scale $k_\theta\rho_s$ of choice. 
If the length scale is chosen appropriately, however, the quasilinear simulation will capture the essential features of the transport mechanism, and it is useful for getting a qualitative understanding of the physical processes.

\section{Simulations} \label{sec:simulations}
%As has been show in previous studies CITERA,\cite{Angioni2006,Nordman2011} the transport of impurities is dependent on the impurity charge. 
%This is also evident from Eq.~\eqref{eq:transport}.
In this paper, the transport of impurities has been studied numerically, by calculating the impurity peaking factor ($PF$) for impurities with various impurity charge ($Z$) and varying values of the driving background gradients. 
The process of calculating the peaking factor is illustrated in Fig.~\ref{fig:Gamma}.
The impurity particle flux $\Gamma_Z$ is computed for $\grad n_Z$ in the vicinity of $\Gamma_Z=0$, taking the estimated residuals of the samples' uncertainty into account (see Fig.~\ref{fig:timetrace}). 
The diffusivity $D_Z$ and convective velocity $RV_Z$ are then given by fitting the acquired fluxes to Eq.~\eqref{eq:transport_short}, where after the peaking factor is obtained as $PF=-\frac{RV_Z}{D_Z}$ (see section~\ref{sec:PF}).

The instabilities causing the transport are fuelled by the free energy present in gradients in the system, and in general the steeper the gradient the more free energy is available, which is expected to lead to stronger modes and more pronounced transport.
Two families of gradients are available that can drive the instabilities: 
the temperature gradients ($-R\grad T_j/T_j\approx R/L_{T_j}$) and the density gradients ($-R\grad n_j/n_j\approx R/L_{n_j}$), where $j=i,\,e$ for main ions and electrons respectively.\footnote{the density and temperature gradients of the impurity species can also drive turbulent transport, however, for trace amounts this effect is negligible}
Numerical studies have been performed, focused on the dependence of the peaking factor on these gradients.

The main parameters used in the simulations are summarised in Tab.~\ref{tab:parameters}.
The parameters where chosen to represent an arbitrary tokamak geometry at about mid radius, and do not represent any one particular experiment.
As can be seen in the table, a TE~or an~ITG~mode dominated plasma was studied by choosing a steep electron temperature gradient ($R/L_{T_Z}=7.0$) together with a moderate ion temperature gradient ($R/L_{T_i}=3.0$) to prompt TE~mode dominated dynamics, and the other way around for ITG~mode dominance. 
It should be noted in this context, that TE~modes can also be driven by steep density gradients.
This case is omitted here, and left for future study.
In order to preserve quasi-neutrality, Eq.~\eqref{eq:qneutrality}, $\grad n_e=\grad n_i$ was used. 
The simulations are limited to cases with $T_e=T_i$.

In order to ensure that the resolution was sufficient, the resolution was varied separately for the perpendicular, parallel and velocity space coordinates, and the effects of this on the mode structure, $k_\perp$ spectra and flux levels were investigated. 
The resolution was then set sufficiently high for the effects on the these indicators to have converged.
For a typical NL simulation for main ions, fully kinetic electrons, and one trace species, a resolution of $n_{x} \times n_{y} \times n_{z} = 96 \times 96 \times 24$ grid points in real space and of $n_{v} \times n_{\mu} = 48 \times 12$ in velocity space was chosen.
For QL~GENE simulations the box size was set to $n_{x} \times n_{y} \times n_{z} = 8 \times 1 \times 24$  and $n_{v} \times n_{\mu} = 64 \times 12$ respectively.

Simulations have been performed with both Deuterons and protons as main ions, but no significant differences in the impurity transport were found between the two cases.

The impurities were included self-consistently as a third species in the simulations, with the trace impurity particle density $n_Z/n_e = 10^{-6}$ in order to ensure that they have a negligible effect on the turbulence.

In the present study, a simple s--$\alpha$ geometry is assumed for the simulation domains.
The effects of different tokamak geometries on drift wave turbulence have been studied in both fluid\cite{Andersson2000,Andersson2002} and gyrokinetic descriptions.\cite{Told2010, Bruckel2010}

\section{Results and discussion} \label{sec:results}
For the scalings studied, the charge number $Z$ of the impurities was varied from $Z=2$ to $Z=74$, with a mass to charge ratio $A/Z=2$.
The scalings of the peaking factor with the temperature gradients were studied by varying $R/L_{T_e}$ between $R/L_{T_e}=6.0$ and $R/L_{T_e}=10.0$ for the TE~mode case, and similarly by varying $R/L_{T_{i,Z}}$ between $R/L_{T_{i,Z}}=6.0$ and $R/L_{T_{i,Z}}=10.0$ for the ITG~mode case.
The density gradient scalings were obtained by varying $R/L_{n_e}$ between $R/L_{n_e}=0.5$ and $R/L_{n_e}=5.0$.

QL and NL scalings of $PF=-RV_Z/D_Z$ were obtained using GENE and compared to results obtained from the fluid model. 

%Common for explaining all results is the balance of the different terms in Eq.~\eqref{eq:transport}.
%Much of the difference between the TE~and ITG~mode dominated cases observed an be understood from convective velocity $V_Z$ in Eq.~\eqref{eq:transport}, which, as elaborated on in section~\ref{sec:PF} above, contains two important terms:\cite{Angioni2006}
% 
% \begin{itemize}
%  \item thermodiffusion:
%  \begin{itemize}
%   \item $V_{T_Z}\sim \frac{1}{Z}\frac{R}{L_{T_Z}}$
%   \item inward for TE $\left(V_{T_Z}<0\right)$, outward for ITG $\left(V_{T_Z}>0\right)$
%  \end{itemize}
%  \item parallel impurity compression: 
%  \begin{itemize}
%   \item $V_{\parallel_Z}\sim\frac{Z}{A_Z}k_\parallel^2\sim\frac{Z}{A_Z q^2}\approx\frac{1}{2 q^2}$
%   \item outward for TE $\left(V_{\parallel_Z}>0\right)$, inward for ITG $\left(V_{\parallel_Z}<0\right)$
%  \end{itemize}
% \end{itemize}
% 
% THE ABOVE SHOULD BE MOVED INTO THE SECTIONS BELOW

\subsection{Scalings with impurity charge} \label{sec:Z}
The $Z$ scalings of the impurity peaking factor for the TE~mode dominated case are presented in Fig.~\ref{fig:Z_TEM}. 
A good agreement between fluid and NL gyrokinetic results is observed for the value $k_\theta\rho_s=0.2$ used in the QL and fluid simulations.
The peaking factors are larger and the trends are more pronounced in the QL~GENE results, which overestimate the peaking factors by approximately a factor of two for low $Z$ impurities.
As expected from the discussion in section~\ref{sec:PF} above, $PF$ varies the most for low $Z$ impurities where the thermopinch is stronger.
For heavier elements, the peaking factor saturates at levels well below neoclassical predictions, as seen in previous gyrokinetic and fluid studies, of both TE~and ITG~mode dominated impurity transport.\cite{Angioni2006,Angioni2007,Nordman2008,Fulop2009,Nordman2011,Angioni2011,Angioni2009a}

For comparison, the results for the ITG~mode dominated case is shown in Fig.~\ref{fig:Z_ITG}.
The two cases show a qualitative difference, with $PF$ falling towards saturation as $Z$ is increased for the TE~mode case, while the opposite holds for the ITG~mode case.
This is in line with previous QL~kinetic and fluid results.\cite{Angioni2006, Angioni2007, Nordman2008, Fulop2009, Nordman2011}
The peaking factor is close to zero for low $Z$ values in the ITG~mode dominated case, however, the sign of $PF$ remains positive for all $Z$ in both the TE~and the ITG~mode dominated case considered. 
This indicates that a net inward pinch is the most common situation in both $TE$~and $ITG$~mode driven impurity transport, for the parameters considered.
It is, however, known from QL as well as NL gyrokinetic simulations that the convection of the impurities can reverse its direction, if the electron heat flux significantly exceeds the ion heat flux.\cite{Angioni2006,Angioni2009a}

\label{sec:Z_TEM_ITG}
The qualitative difference between the $Z$ scalings for the TE~and ITG~mode dominated cases can be understood from the balance of the thermodiffusion and parallel impurity compression in Eq.~\eqref{eq:transport}, the two terms having opposite signs for TE~and ITG, as discussed above (section~\ref{sec:PF}).
The parallel impurity compression is almost independent of $Z$, so it can be assumed that the thermodiffusion is the main contributor to the observed trends.
The thermodiffusion, on the other hand, has the strongest effect for low $Z$ values, explaining the drop and rise of $PF$ with $Z$ for the TE~and the ITG~mode respectively.
Since this term goes to zero for large values of $Z$, this also explains the observed saturation.

\label{sec:kpar}
In the fluid treatment, a strong ballooning eigenfunction is assumed with $k_{\parallel,sb}^2 = \left(3q^2 R^2\right)^{-1}$.\cite{Hirose1994}
Since the contribution from the parallel compression pinch depends on the mode structure along the field line, the results are expected to be sensitive to the choice of $k_\parallel$. 
To investigate the sensitivity of the fluid results to the mode structure, a simplified treatment was used, varying $k_\parallel$ around its strong ballooning value while keeping the eigenvalues fixed.
The results are shown in Fig.~\ref{fig:kpar} for $k_\theta\rho_s=0.2$ and~$0.3$ in the TE~and ITG~mode dominated cases. 
As observed, the peaking factors for TE~mode turbulence is sensitive to the choice of $k_\parallel$, with the peaking factor going from $PF\approx2$ to $PF\approx0$ when $k_\parallel^2$ is varied from $0.5$ to $2$ times its strong ballooning value.
%The relative contribution of the parallel compression term to the convective flux will thus depend on $k_\parallel$, and as a consequence the peaking factors will be affected.

As is evident from Fig.~\ref{fig:Z}, the value of $PF$ is also dependent on the choice of $k_\theta\rho_s$, the perpendicular length scale. 
Finding the $k_\theta\rho_s$ that allows the QL~gyrokinetic and fluid models to best capture the behaviour of the impurity transport is non-trivial.
For the cases considered, the results were obtained with $k_\theta\rho_s=0.2$.
This is in line with previous results regarding comparisons of  fluid and NL gyrokinetic results.\cite{Nordman2007a}
The nonlinear spectra for the fluctuations in the background electrostatic potential ($\phi$) are illustrated in Fig.~\ref{fig:k_spectra} for the TE~and ITG~mode dominated cases in Fig.~\ref{fig:Z}.

The spectra both show a peak in the fluctuations at $k_\theta\rho_s \approx 0.15$, well below the wave number of maximum linear growth rate, $k_\theta\rho_s\approx0.3$.
We have confirmed that for $k_\theta\rho_s$ in the range $0.15$--$0.4$, qualitatively similar QL~results are obtained.
In the following, $k_\theta\rho_s=0.2$--$0.3$ will be used.

A further complication that arises when studying TE~mode turbulence is the onset of electron temperature gradient (ETG) driven modes.
These are mostly sub-dominant, and so are not captured by the QL~treatment, but may give a nonlinear contribution through the nonlinear coupling between the small scale ETG modes and the longer wave lengths of the dominant TE modes, and care has to be taken to avoid this effect.\cite{Ernst2009, Gao2005}

\subsection{Scalings with the temperature gradients} \label{sec:omt}
The obtained scalings of $PF$ with the electron temperature gradient are presented in Fig.~\ref{fig:omt_TEM}. 
We note that the QL gyrokinetic simulations overestimate the peaking factors by up to $\sim50\%$.
The fluid results are in good agreement with the NL~GENE results.
Only weak trends were observed, in compliance with previous studies.\cite{Nordman2007a, Nordman2007b, Fulop2009}
%The NL simulations yield similiar results, both quantitatively and qualitatively, to both the QL and fluid results for the lengthscales chosen.
As with the $Z$ scaling in Fig.~\ref{fig:Z_TEM}, the NL trend is less pronounced, reaching saturation for lower values of $R/L_{T_e}$ than the other two models.

For comparison, the results for the ITG~mode dominated case are shown in Fig.~\ref{fig:omt_ITG}.
As was observed for the $Z$ scaling in section~\ref{sec:Z} above, the trends for the TE~and ITG~mode dominated case are reversed; $PF$ rises with driving gradient for the TE~case, but falls for the ITG~case.
The difference between the two trends can be understood in part from the thermodiffusion in Eq.~\eqref{eq:transport}.
This term grows more important as the ion/impurity temperature gradient steepens, providing a strong outward pinch for the ITG~mode dominated impurity transport and thus yielding lower values of $PF$ as $R/L_{T_Z}$ increases (Fig.~\ref{fig:omt_ITG}).
Since the impurity temperature gradient is constant for the $\grad T_e$ scaling, however, other effects are behind the TE~mode scaling in Fig.~\ref{fig:omt_TEM}.
The eigenvalues, in particular the mode growth rates, grow with $\grad T_{e,i}$, as shown in Fig.~\ref{fig:omt_eigens}.
This will alter the relative contributions of the convective terms in Eq.~\ref{eq:transport}, and hence affect the peaking factor.
We note here that the eigenvalues in Fig.~\ref{fig:omt_eigens} are normalised to $c_s/R$, giving $\omega_r<0$ for TE~modes and $\omega_r>0$ for ITG~modes.

As with the $Z$ scaling, the sign of $PF$ usually remains positive for the $\grad T_{e,i}$ scalings, though a modest flux reversal is observed when the trends of the scalings with $Z$ and $R/L_{T_i}$ for the ITG~mode 
combine. 
This is the case for He in Fig.~\ref{fig:omt_ITG}.
The flux reversal is observed only for very steep temperature gradients for the considered parameter values with $T_e=T_i$.

\subsection{Scalings with density gradient} \label{sec:omn}
In experimentally relevant situations where the impurity and main ion fuelling originates from the edge, the core impurity and background density peaking factors should be calculated self-consistently for zero particle flux.
For this purpose, the equations $\Gamma_Z=0$ and $\Gamma_{i,e}=0$ need to be solved self-consistently.
This is in the following achieved by varying the main ion density gradient $R/L_{n_e}$ until $\Gamma_e=0$ is obtained, and using the zero flux background density gradient in the impurity transport calculations.
We assume trace levels of impurities and use the fluid model for simplicity.
The results are illustrated in Fig.~\ref{fig:omn} which shows the impurity peaking factor $R/L_{n_z}$ versus $R/L_{n_e}$ for both the TE~and ITG~mode dominated cases.
The value of $R/L_{n_e}$ for zero background particle flux is marked in the figure.
We note that the background density peaking is larger than the impurity peaking with $R/L_{n_e}=3.0$ for the TE~case and $R/L_{n_e}=2.5$ for the ITG~case.
We emphasise that the result is obtained using a collision-less model.
It is known that collisions have a large impact on the background density peaking in both fluid\cite{Nordman2011} and gyrokinetic models.\cite{Angioni2009b}

For the $R/L_{n_e}$ scaling, the same trends are observed in both GENE and fluid data, with a strong sensitivity for lower $Z$ impurities.
This is particularly evident for the ITG~mode case in Fig.~\ref{fig:omn_ITG}, where the peaking factor for the He impurity shows a marked increase as $\grad n_e$ steepens for both GENE and fluid results, whereas for the heavier elements a nearly flat dependence is observed.

As shown in Fig.~\ref{fig:omn_eigens}, the eigenvalues vary with the electron density gradient.
A reduction of $|\omega_r|$ and an increase of $\gamma$ are observed with increasing $R/L_{n_e}$, which leads to a reduction of the relative amplitude of the thermopinch in Eq.~\eqref{eq:transport}.
This explains the observed $PF$ scaling for the TE~and ITG~mode driven cases in Fig.~\ref{fig:omn_TEM} and Fig.~\ref{fig:omn_ITG} respectively.

As with the $\grad T_i$ scaling, the combined effect of the $Z$ and $\grad n_e$ scalings is observed to lead to a flux reversal for the He impurity in the ITG~mode dominated case in Fig.~\ref{fig:omn_ITG}.
This happens for flat electron density profiles in the QL GENE results. 
Outside of this regime the sign of $PF$ remains positive.

%\section{Discussion} \label{sec:discussion}
%LET THERE BE NO DISCUSSION!

\section{Conclusion and future work} \label{sec:conclusions}
In the present paper the transport of impurities driven by trapped electron (TE)~mode driven turbulence has been studied. 
Non-linear~(NL) gyrokinetic simulations using the code GENE were compared with results from quasilinear~(QL) gyrokinetic simulations and a computationally efficient fluid model, viable for use in predictive simulations.
The main focus has been on model comparisons for electron temperature gradient driven turbulence regarding the sign of the convective impurity velocity~(pinch) and the impurity peaking factor $\left(R/L_{n_Z}\right)$ for zero impurity flux.
In particular, the scaling of the peaking factor with impurity charge $Z$ and with driving temperature gradient has been investigated and compared with the more well known results for Ion Temperature Gradient (ITG) driven turbulence.

For the scaling of the peaking factor with the impurity charge $Z$, a weak dependence was obtained from NL~GENE simulations, which was reproduced well by the fluid simulations.
The QL GENE results showed a stronger dependence for low $Z$ impurities and overestimated the peaking factor by up to a factor of two in this region.
As in the case of ITG dominated turbulence, the peaking factors were found to saturate as $Z$ increased, at a level much below neoclassical predictions.
However, the scaling with $Z$ was found to be weak or reversed as compared to the ITG case, where the larger peaking factors were obtained for high $Z$ impurities.

Using the fluid model it was shown that the impurity peaking factors in the TE~mode dominate case are sensitive to the mode structure along the field lines ($k_\parallel$) through the parallel compression pinch.
It was shown that assuming a strong ballooning eigenfunction with $k_\parallel^2=\left(3 q^2 R^2\right)^{-1}$, together with $k_\theta\rho_s=0.2$, gave a good agreement with the results from the NL~GENE simulations.

The scaling of impurity peaking with the driving background temperature gradients were found to be weak in most cases. 
The QL results were also here found to significantly overestimate the peaking factor for low $Z$ values.

The main ion peaking relative to the impurity peaking was studied using a self-consistent treatment of the main ion and impurity particle fluxes.
It was found that the main ion peaking was slightly larger than the impurity peaking, for both TE~and ITG~mode dominated turbulence.
These results were obtained using the fluid model in the collision-less limit.

The present study is based on low $\beta$ plasmas in a simple s--$\alpha$ circular tokamak equilibrium.
Future work will aim to study the effects of more realistic geometries, finite $\beta$, as well as effects of plasma rotation on impurity transport in NL fluid and gyrokinetic descriptions.

\section{Acknowledgements}
The simulations were performed on resources provided on the Lindgren\cite{Lindgren} and HPC-FF\cite{HPC-FF} high performance computers, by the
Swedish National Infrastructure for Computing (SNIC) at Paralleldatorcentrum
(PDC) and the European Fusion Development Agreement (EFDA), respectively.

J. Vincent at PDC and T. Görler at IPP-Garching are acknowledged for their
assistance concerning technical and implementational aspects in making the GENE
code run on the PDC Lindgren super-computer.

A. Strand and L. Strand at Herrgårdsskolan are acknowledged for their help with the nonlinear simulations.

The authors would also like to thank F. Jenko, M. J. Püschel, F. Merz and the rest of the GENE team at IPP-Garching for their valuable support and input.

\newpage 
\bibliographystyle{unsrtnat} % fort lic
\bibliography{fusion.bib}

\begin{thebibliography}{50}
\providecommand{\natexlab}[1]{#1}
\providecommand{\url}[1]{\texttt{#1}}
\expandafter\ifx\csname urlstyle\endcsname\relax
  \providecommand{\doi}[1]{doi: #1}\else
  \providecommand{\doi}{doi: \begingroup \urlstyle{rm}\Url}\fi

\bibitem[Harte et~al.(2010)Harte, Suzuki, and {Kato, et al}]{Harte2010}
C.~S. Harte, C.~Suzuki, and T.~{Kato, et al}.
\newblock \emph{J. Phys. B}, 43\penalty0 (20):\penalty0 205004, 2010.

\bibitem[Fr\"ojdh et~al.(1992)Fr\"ojdh, Liljestr\"om, and Nordman]{Frojdh1992}
M.~Fr\"ojdh, M.~Liljestr\"om, and H.~Nordman.
\newblock \emph{Nucl. Fusion}, 32\penalty0 (3):\penalty0 419,   1992.

\bibitem[Basu et~al.(2003)Basu, Jessen, Naulin, and Rasmussen]{Basu2003}
R.~Basu, T.~Jessen, V.~Naulin, and J.~Juul Rasmussen.
\newblock \emph{Phys. Plasmas}, 10\penalty0 (7):\penalty0 2696,   2003.

\bibitem[Estrada-Mila et~al.(2005)Estrada-Mila, Candy, and
  Waltz]{Estrada-Mila2005}
C.~Estrada-Mila, J.~Candy, and R.W. Waltz.
\newblock \emph{Phys. Plasmas}, 12\penalty0 (2):\penalty0 022305,   2005.

\bibitem[Naulin(2005)]{Naulin2005}
V.~Naulin.
\newblock \emph{Phys. Rev. E}, 71\penalty0 (1):\penalty0 015402,   2005.
\newblock \doi{10.1103/PhysRevE.71.015402}.

\bibitem[Priego et~al.(2005)Priego, Garcia, Naulin, and Rasmussen]{Priego2005}
M.~Priego, O.~E. Garcia, V.~Naulin, and J.~Juul Rasmussen.
\newblock \emph{Phys. Plasmas}, 12\penalty0 (6):\penalty0 062312,   2005.

\bibitem[F\"ul\"op and Weiland(2006)]{Fulop2006}
T.~F\"ul\"op and J.~Weiland.
\newblock \emph{Phys. Plasmas}, 13\penalty0 (11):\penalty0 112504,   2006.

\bibitem[Bourdelle et~al.(2007)Bourdelle, Garbet, Imbeaux, Casati, Dubuit,
  Guirlet, and Parisot]{Bourdelle2007}
C.~Bourdelle, X.~Garbet, F.~Imbeaux, A.~Casati, N.~Dubuit, R.~Guirlet, and
  T.~Parisot.
\newblock \emph{Phys. Plasmas}, 14\penalty0 (11):\penalty0 112501,   2007.

\bibitem[Dubuit et~al.(2007)Dubuit, Garbet, Parisot, Guirlet, and
  Bourdelle]{Dubuit2007}
N.~Dubuit, X.~Garbet, T.~Parisot, R.~Guirlet, and C.~Bourdelle.
\newblock \emph{Phys. Plasmas}, 14\penalty0 (4):\penalty0 042301,   2007.

\bibitem[Camenen et~al.(2009)Camenen, Peeters, Angioni, Casson, Hornsby,
  Snodin, and Strintzi]{Camenen2009}
Y.~Camenen, A.~G. Peeters, C.~Angioni, F.~J Casson, W.~A Hornsby, A.~P. Snodin,
  and D.~Strintzi.
\newblock \emph{Phys. Plasmas}, 16\penalty0 (1):\penalty0 012503,   2009.

\bibitem[F\"ul\"op et~al.(2010)F\"ul\"op, Braun, and Pusztai]{Fulop2010}
T.~F\"ul\"op, S.~Braun, and I.~Pusztai.
\newblock \emph{Phys. Plasmas}, 17\penalty0 (6):\penalty0 062501,   2010.

\bibitem[Futatani et~al.(2010)Futatani, Garbet, Benkadda, and
  Dubuit]{Futatani2010}
S.~Futatani, X.~Garbet, S.~Benkadda, and N.~Dubuit.
\newblock \emph{Phys. Rev. Lett.}, 104\penalty0 (1):\penalty0 015003,   2010.

\bibitem[Hein and Angioni(2010)]{Hein2010}
T.~Hein and C.~Angioni.
\newblock \emph{Phys. Plasmas}, 17\penalty0 (1):\penalty0 012307,   2010.

\bibitem[Moradi et~al.(2010)Moradi, Tokar, and Weyssow]{Moradi2010}
S.~Moradi, M.~Z. Tokar, and B.~Weyssow.
\newblock \emph{Phys. Plasmas}, 17\penalty0 (1):\penalty0 012101,   2010.

\bibitem[F\"ul\"op and Moradi(2011)]{Fulop2011}
T.~F\"ul\"op and S.~Moradi.
\newblock \emph{Phys. Plasmas}, 18\penalty0 (3):\penalty0 030703,   2011.

\bibitem[Nordman et~al.(2008)Nordman, Singh, and et~al.]{Nordman2008}
H.~Nordman, R.~Singh, and T.~F\"ul\"op et~al.
\newblock \emph{Phys. Plasmas}, 15:\penalty0 042316, 2008.

\bibitem[Angioni and Peeters(2006)]{Angioni2006}
C.~Angioni and A.~G. Peeters.
\newblock \emph{Phys. Rev. Lett.}, 96:\penalty0 095003,   2006.

\bibitem[et~al.(2007)]{Angioni2007}
C.~Angioni et~al.
\newblock \emph{Phys. Plasmas}, 14\penalty0 (5):\penalty0 055905,   2007.

\bibitem[Nordman et~al.(2007{\natexlab{a}})Nordman, F\"ul\"op, Candy, Strand,
  and Weiland]{Nordman2007a}
H.~Nordman, T.~F\"ul\"op, J.~Candy, P.~Strand, and J.~Weiland.
\newblock \emph{Phys. Plasmas}, 14\penalty0 (5):\penalty0 052303,
  2007{\natexlab{a}}.

\bibitem[Angioni et~al.(2009{\natexlab{a}})Angioni, Peeters, Pereverzev,
  Bottino, Candy, Dux, Fable, Hein, and Waltz]{Angioni2009a}
C.~Angioni, A.~G. Peeters, G.~V. Pereverzev, A.~Bottino, J.~Candy, R.~Dux,
  E.~Fable, T.~Hein, and R.~E. Waltz.
\newblock \emph{Nucl. Fusion}, 49\penalty0 (5):\penalty0 055013,
  2009{\natexlab{a}}.

\bibitem[F\"ul\"op and Nordman(2009)]{Fulop2009}
T.~F\"ul\"op and H.~Nordman.
\newblock \emph{Phys. Plasmas}, 16\penalty0 (3):\penalty0 032306,   2009.

\bibitem[Nordman et~al.(2011)Nordman, Skyman, Strand, Giroud, Jenko, and
  et~al.]{Nordman2011}
H.~Nordman, A.~Skyman, P.~Strand, C.~Giroud, F.~Jenko, and F.~Merz et~al.
\newblock \emph{Plasma Phys. Contr. F.}, 53\penalty0 (10):\penalty0 105005,
  2011.

\bibitem[Dux et~al.(2003)Dux, Neu, Peeters, Pereverzev, Mück, Ryter, and
  Stober]{Dux2003}
R.~Dux, R.~Neu, A.~G. Peeters, G.~Pereverzev, A.~Mück, F.~Ryter, and
  J.~Stober.
\newblock \emph{Plasma Phys. Contr. F.}, 45\penalty0 (9):\penalty0 1815,
  2003.

\bibitem[Puiatti et~al.(2003)Puiatti, Valisa, and {Mattioli, et
  al}]{Puiatti2003}
M.~E. Puiatti, M.~Valisa, and M.~{Mattioli, et al}.
\newblock \emph{Plasma Phys. Contr. F.}, 45\penalty0 (12):\penalty0 2011,
  2003.

\bibitem[Puiatti et~al.(2006)Puiatti, Valisa, and {Angioni, et
  al}]{Puiatti2006}
M.~E. Puiatti, M.~Valisa, and C.~{Angioni, et al}.
\newblock \emph{Phys. Plasmas}, 13\penalty0 (4):\penalty0 042501,   2006.

\bibitem[Giroud et~al.(2007)Giroud, Barnsley, and {Buratti, et al}]{Giroud2007}
C.~Giroud, R.~Barnsley, and P.~{Buratti, et al}.
\newblock \emph{Nucl. Fusion}, 47\penalty0 (4):\penalty0 313,   2007.

\bibitem[{Giroud et al}(2009)]{Giroud2009}
C.~{Giroud et al}.
\newblock   2009.
\newblock Princeton, USA.

\bibitem[Helander and Digmar(2002)]{Helander2002}
P.~Helander and D.~J. Digmar.
\newblock \emph{Collisional Transport in Magnetized Plasmas}.
\newblock Cambridge University Press, 2002.

\bibitem[Angioni et~al.(2011)Angioni, McDermott, Fable, Fischer, Pütterich,
  Ryter, Tardini, and {the ASDEX Upgrade Team}]{Angioni2011}
C.~Angioni, R.M. McDermott, E.~Fable, R.~Fischer, T.~Pütterich, F.~Ryter,
  G.~Tardini, and {the ASDEX Upgrade Team}.
\newblock \emph{Nucl. Fusion}, 51\penalty0 (2):\penalty0 023006, 2011.

\bibitem[Jenko et~al.(2000)Jenko, Dorland, Kotschenreuther, and
  Rogers]{Jenko2000}
F.~Jenko, W.~Dorland, M.~Kotschenreuther, and B.~N. Rogers.
\newblock \emph{Phys. Plasmas}, 7\penalty0 (5):\penalty0 1904,   2000.

\bibitem[Dannert(2005)]{Dannert2005a}
T.~Dannert.
\newblock \emph{Gyrokinetische {S}imulation von {P}lasmaturbulenz mit
  gefangenen {T}eilchen und elektromagnetischen {E}ffekten}.
\newblock Ph.d. thesis (monography), Technischen {U}niversit\"at {M}\"unchen,
  2005.

\bibitem[Merz(2008)]{Merz2008a}
F.~Merz.
\newblock \emph{Gyrokinetic Simulation of Multimode Plasma Turbulence}.
\newblock Ph.d. thesis (monography), Westf\"alischen Wilhelms-Universit\"at
  M\"unster, 2008.

\bibitem[Weiland(2000)]{Weiland2000}
J.~Weiland.
\newblock \emph{Collective Modes in Inhomogeneous Plasmas}.
\newblock IoP Publishing, 2000.

\bibitem[Nordman et~al.(2007{\natexlab{b}})Nordman, Strand, and
  Garbet]{Nordman2007b}
H.~Nordman, P.~Strand, and X.~Garbet.
\newblock \emph{J. Plasma Phys.}, 73\penalty0 (5):\penalty0 731--740,
  2007{\natexlab{b}}.

\bibitem[Hirose et~al.(1994)Hirose, Zhang, and Elia]{Hirose1994}
A.~Hirose, L.~Zhang, and E.~Elia.
\newblock \emph{Phys. Rev. Lett.}, 72\penalty0 (25):\penalty0 3993--3996,
  1994.

\bibitem[GEN(2011)]{GENE}
  2011.
\newblock URL \url{http://www.ipp.mpg.de/~fsj/gene/}.

\bibitem[Antonsen and Lane(1980)]{Antonsen1980}
T.~M. Antonsen and B.~Lane.
\newblock \emph{Phys. Fluids}, 23\penalty0 (6):\penalty0 1205,   1980.

\bibitem[Frieman and Chen(1982)]{Frieman1982}
E.~A. Frieman and L.~Chen.
\newblock \emph{Phys. Fluids}, 25\penalty0 (3):\penalty0 502,   1982.

\bibitem[Hahm et~al.(1988)Hahm, Lee, and Brizard]{Hahm1988}
T.~S. Hahm, W.~W. Lee, and A.~Brizard.
\newblock \emph{Phys. Fluids}, 31\penalty0 (7):\penalty0 1940,   1988.

\bibitem[Brizard(1989)]{Brizard1989}
A.~Brizard.
\newblock \emph{J. Plasma Phys.}, 41\penalty0 (3):\penalty0 541,   1989.

\bibitem[Dannert and Jenko(2005)]{Dannert2005b}
T.~Dannert and F.~Jenko.
\newblock \emph{Phys. Plasmas}, 12\penalty0 (7):\penalty0 072309,   2005.

\bibitem[Anderson et~al.(2000)Anderson, Nordman, and Weiland]{Andersson2000}
J.~Anderson, H.~Nordman, and J.~Weiland.
\newblock \emph{Plasma Phys. Contr. F.}, 42\penalty0 (5):\penalty0 545,   2000.

\bibitem[Anderson et~al.(2002)Anderson, Rafiq, Nadeem, and
  Persson]{Andersson2002}
J.~Anderson, T.~Rafiq, M.~Nadeem, and M.~Persson.
\newblock \emph{Phys. Plasmas}, 9\penalty0 (5):\penalty0 1629,   2002.

\bibitem[Told and Jenko(2010)]{Told2010}
D.~Told and F.~Jenko.
\newblock \emph{Phys. Plasmas}, 17\penalty0 (4):\penalty0 042302,   2010.

\bibitem[Bruckel et~al.(2010)Bruckel, Sauter, Angioni, Candy, Fable, and
  Lapillonne]{Bruckel2010}
A.~Bruckel, O.~Sauter, C.~Angioni, J.~Candy, E.~Fable, and X.~Lapillonne.
\newblock \emph{J. Phys.: Conf. Ser.}, 260\penalty0 (1):\penalty0 012006, 2010.

\bibitem[Ernst et~al.(2009)Ernst, Lang, Nevins, Hoffman, and Chen]{Ernst2009}
D.~R. Ernst, J.~Lang, W.~M. Nevins, M.~Hoffman, and Y.~Chen.
\newblock \emph{Phys. Plasmas}, 16\penalty0 (5):\penalty0 055906,   2009.

\bibitem[Gao et~al.(2005)Gao, Sanuki, Itoh, and Dong]{Gao2005}
Z.~Gao, H.~Sanuki, K.~Itoh, and J.~Q. Dong.
\newblock \emph{Phys. Plasmas}, 12\penalty0 (2):\penalty0 022503,   2005.

\bibitem[Angioni et~al.(2009{\natexlab{b}})Angioni, Candy, Fable, Maslov,
  Peeters, Waltz, and Weisen]{Angioni2009b}
C.~Angioni, J.~Candy, E.~Fable, M.~Maslov, A.~G. Peeters, R.~E. Waltz, and
  H.~Weisen.
\newblock \emph{Phys. Plasmas}, 16\penalty0 (6):\penalty0 060702,
  2009{\natexlab{b}}.

\bibitem[Lin()]{Lindgren}
URL \url{http://www.pdc.kth.se/resources/computers/lindgren/}.

\bibitem[HPC()]{HPC-FF}
URL \url{http://www2.fz-juelich.de/jsc/juropa/}.

\end{thebibliography}

%
%
% Figures:
%
%
\newpage

\begin{table}[\figplacing]
 \centering
% \scriptsize
 \caption[Parameters]{\small Parameters used in the gyrokinetic simulations, \note{\dag} denotes scan parameters}
 \label{tab:parameters}
 \begin{tabular}{l||r|r}\hline
 & ITG: & TE: \\ \hline\hline
 $T_i/T_e$:             & $1.0$     & $1.0$     \\
 $s$:             & $0.8$     & $0.8$     \\
 $q$:                 & $1.4$     & $1.4$     \\
 $\epsilon=r/R$:        & $0.14$    & $0.14$    \\
 $n_e$, $n_i+n_Z$:      & $1.0$     & $1.0$     \\ %& $10^{19}$\unit{m^{-3}}\\
 $n_Z$ \emph{(trace)}:  & $10^{-6}$ & $10^{-6}$ \\
% $R$:                   & $1.0$   & $1.0$\\ %&\unit{m}\\
 $R/L_{n_{i,e}}$:\note{\dag}       & $2.0$--$3.0$   & $2.0$--$3.0$ \\
 $R/L_{T_i},R/L_{T_Z}$:\note{\dag} & $7.0$   & $3.0$ \\
 $R/L_{T_e}$:\note{\dag}           & $3.0$   & $7.0$ \\ %\hline
% $\left(l_x,\, l_y\right)$:  & $\left(2\pi,\, 2\pi/k\rho\right)\cdot\rho$ & $\left(2\pi,\, 2\pi/k\rho\right)\cdot\rho$ \\
% $N_x\times N_{ky}\times N_z$: & $8\times1\times24$ & $8\times1\times24$\\
% $N_{v_{||}}\times N_\mu$:     & $64\times12$ & $64\times12$
 \end{tabular}
\end{table}

\clearpage

\clearpage

\begin{figure}[\figplacing] %fig:Z_TEM
 \centering
 \subfloat[dependence of the peaking factor ($PF$) on $Z$ for the TE~case \label{fig:Z_TEM}] 
 {\includegraphics[width=\figwidth]{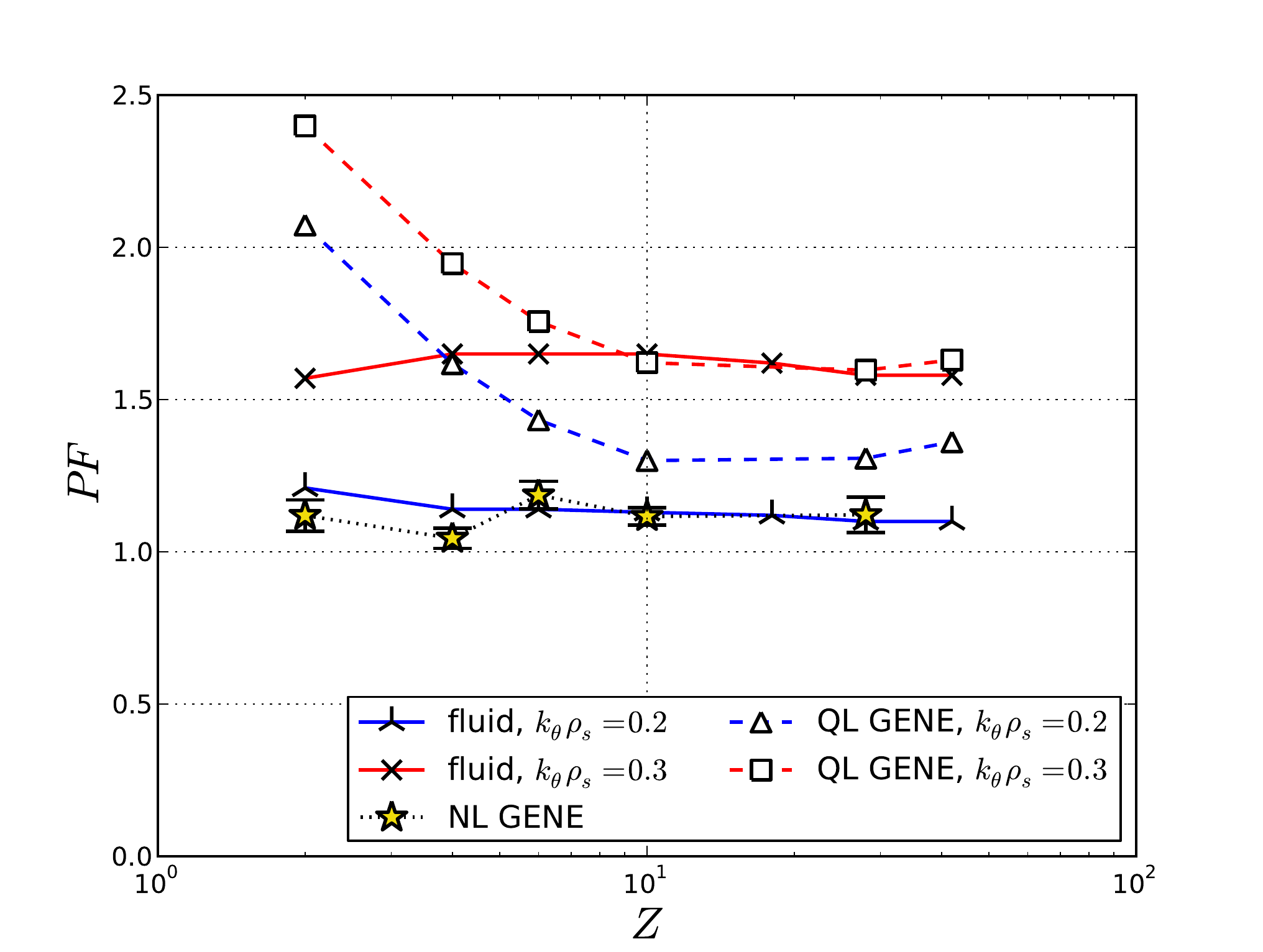}}
 
\phantomcaption{}
\end{figure}

\begin{figure} %fig:Z_ITG, fig:Z
 \ContinuedFloat
 \centering
 \subfloat[dependence of the peaking factor ($PF$) on $Z$ for the ITG~case \label{fig:Z_ITG}] 
 {\includegraphics[width=\figwidth]{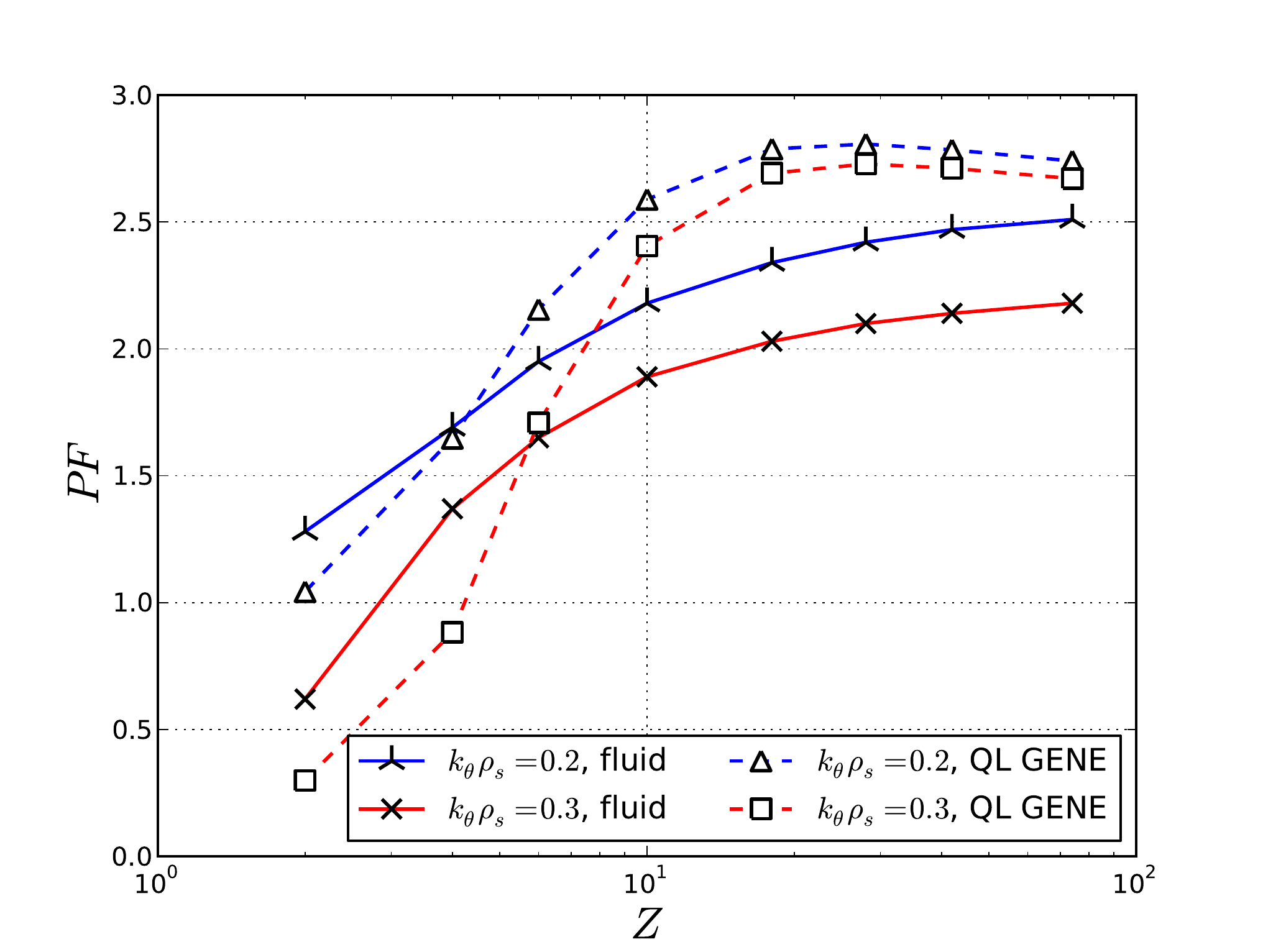}}
 
 \caption{Scalings of the peaking factor ($PF$) with impurity charge ($Z$). 
 Parameters are $q=1.4$, $s=0.8$, $\epsilon=r/R=0.143$ in both subfigures, with $R/L_{T_i}=R/L_{T_Z}=3.0$, $R/L_{T_e}=7.0$, $R/L_{n_e}=2.0$ for the TE~case~(Fig.~\ref{fig:Z_TEM}), and $R/L_{T_i}=R/L_{T_Z}=7.0$, $R/L_{T_e}=3.0$, $R/L_{n_e}=3.0$ for the ITG~case~(Fig.~\ref{fig:Z_ITG}).
 %The main ion species is H in the former case, D in the latter.
 The error bars for the NL~GENE results in Fig.~\ref{fig:Z_TEM} indicate an estimated error of one standard deviation.}
 \label{fig:Z}
\end{figure}

\clearpage

\begin{figure}[\figplacing] %fig:kpar
 \centering
 \includegraphics[width=\figwidth]{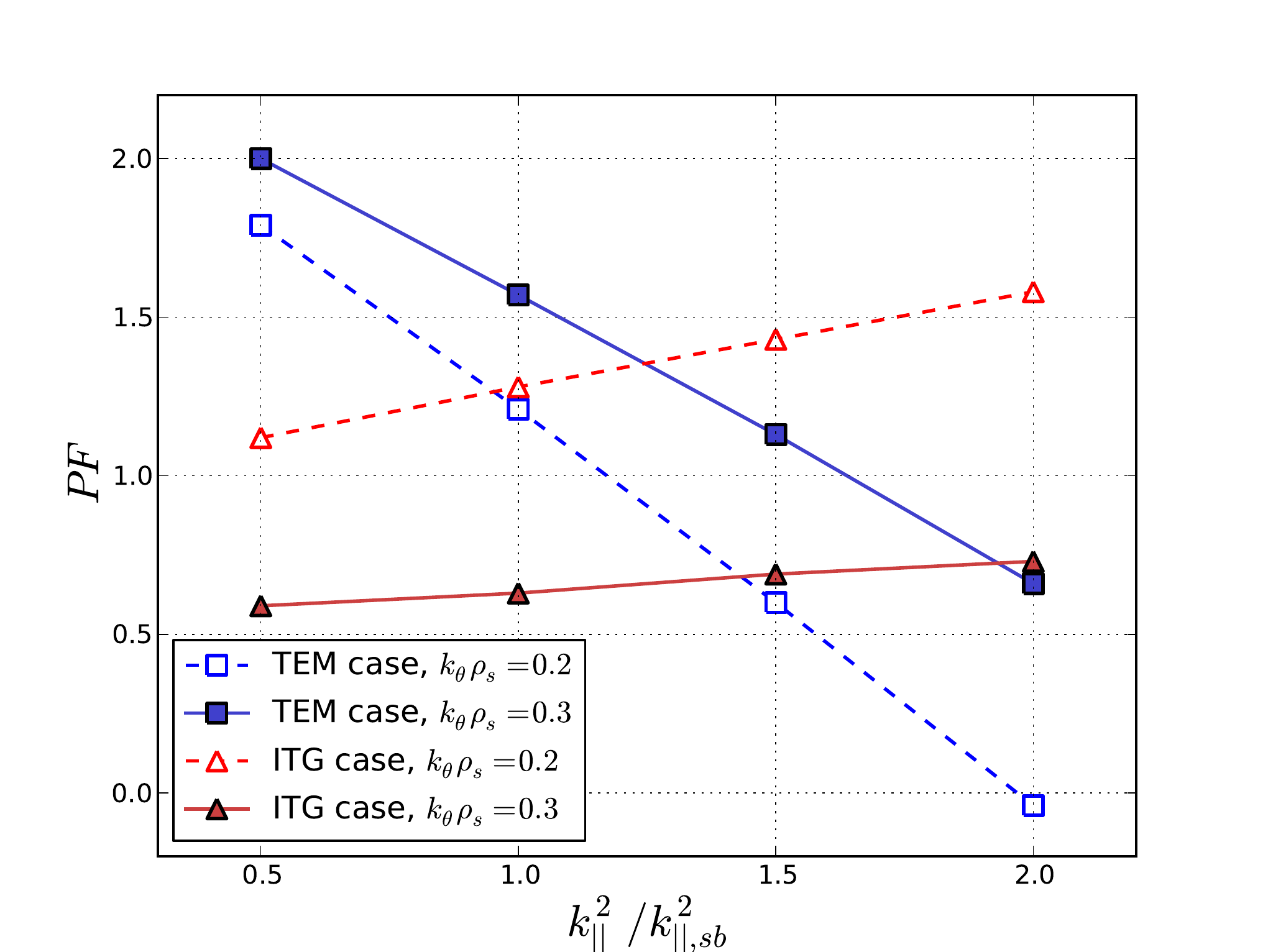}
 \caption{Scaling of the peaking factor ($PF$) with $k_\parallel^2/k_{\parallel,sb}^2$ for He~impurity, where $k_{\parallel,sb}^2=\left(3q^2R^2\right)^{-1}$ is the strong ballooning value;\cite{Hirose1994} fluid results with parameters as in Fig.~\ref{fig:Z_TEM}~(TE) and Fig.~\ref{fig:Z_ITG}~(ITG).}
 \label{fig:kpar}
\end{figure}

\begin{figure}[\figplacing] %fig:k_spectra
 \centering
 \includegraphics[width=\figwidth]{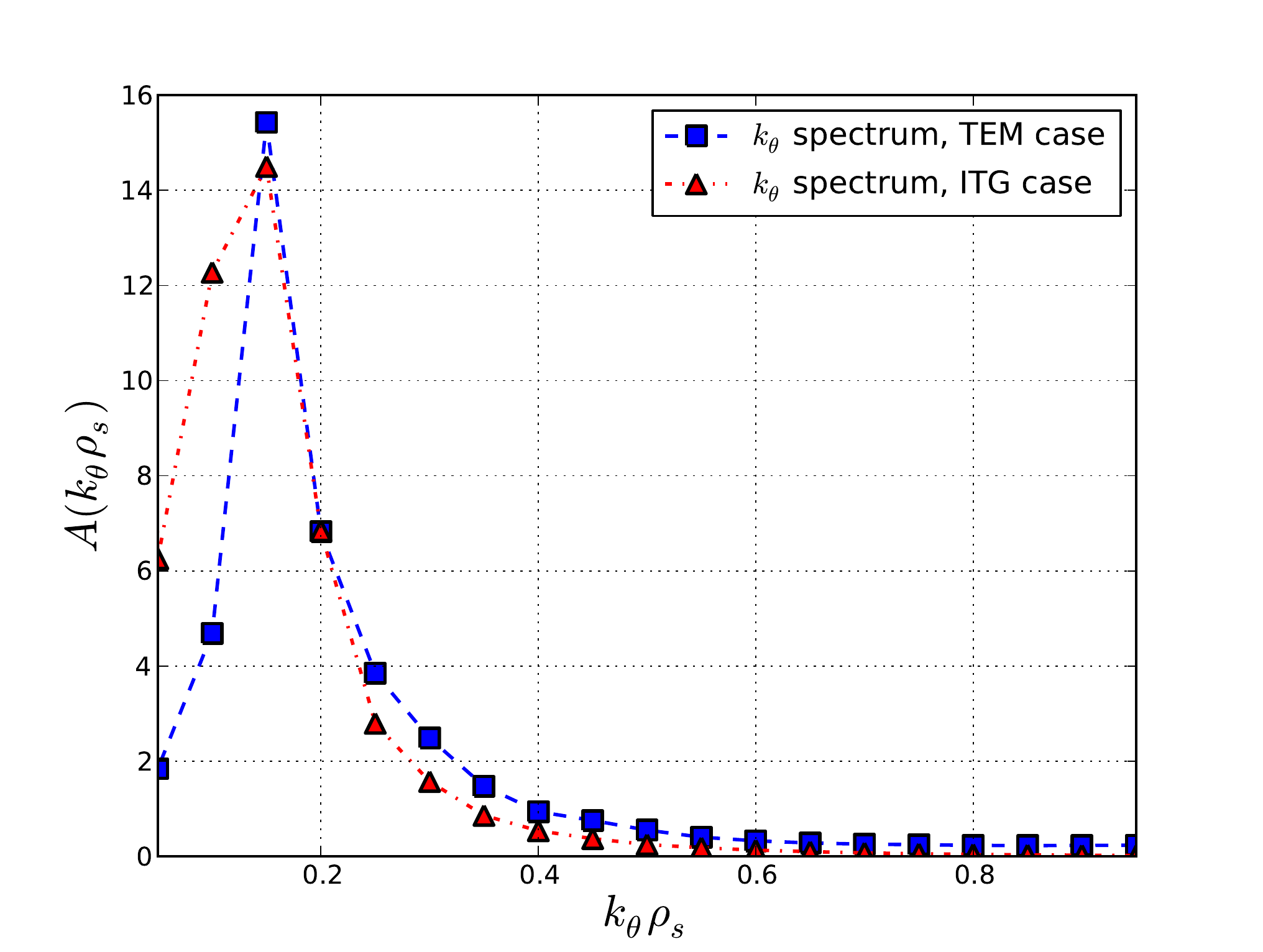}
 \caption{Spectra showing the normalised amplitude ($A(k_\theta\rho_s)$) of the fluctuations in the background electrostatic potential ($\phi$) as a function of $k_\theta\rho_s$; NL~GENE results with parameters as in Fig.~\ref{fig:Z_TEM} (TE) and Fig.~\ref{fig:Z_ITG}~(ITG).}
 \label{fig:k_spectra}
\end{figure}

% \begin{figure}[\figplacing] %fig:Dz_RVz_TEM
%  \centering
%  \includegraphics[width=\figwidth]{Dz_RVz_TEM}
%  \caption{Impurity diffusion ($D_Z$) and pinch ($RV_Z$) for the NL data and fluid data for $k_\theta\rho_s=0.2$ in Fig.~\ref{fig:Z_TEM}.}
%  \label{fig:Dz_RVz_TEM}
% \end{figure}

\clearpage

\begin{figure}[\figplacing] %fig:omt_TEM
 \centering
 \subfloat[dependence of the peaking factor ($PF$) on the normalised electron temperature gradient for the TE~case \label{fig:omt_TEM}]
 {\includegraphics[width=\figwidth]{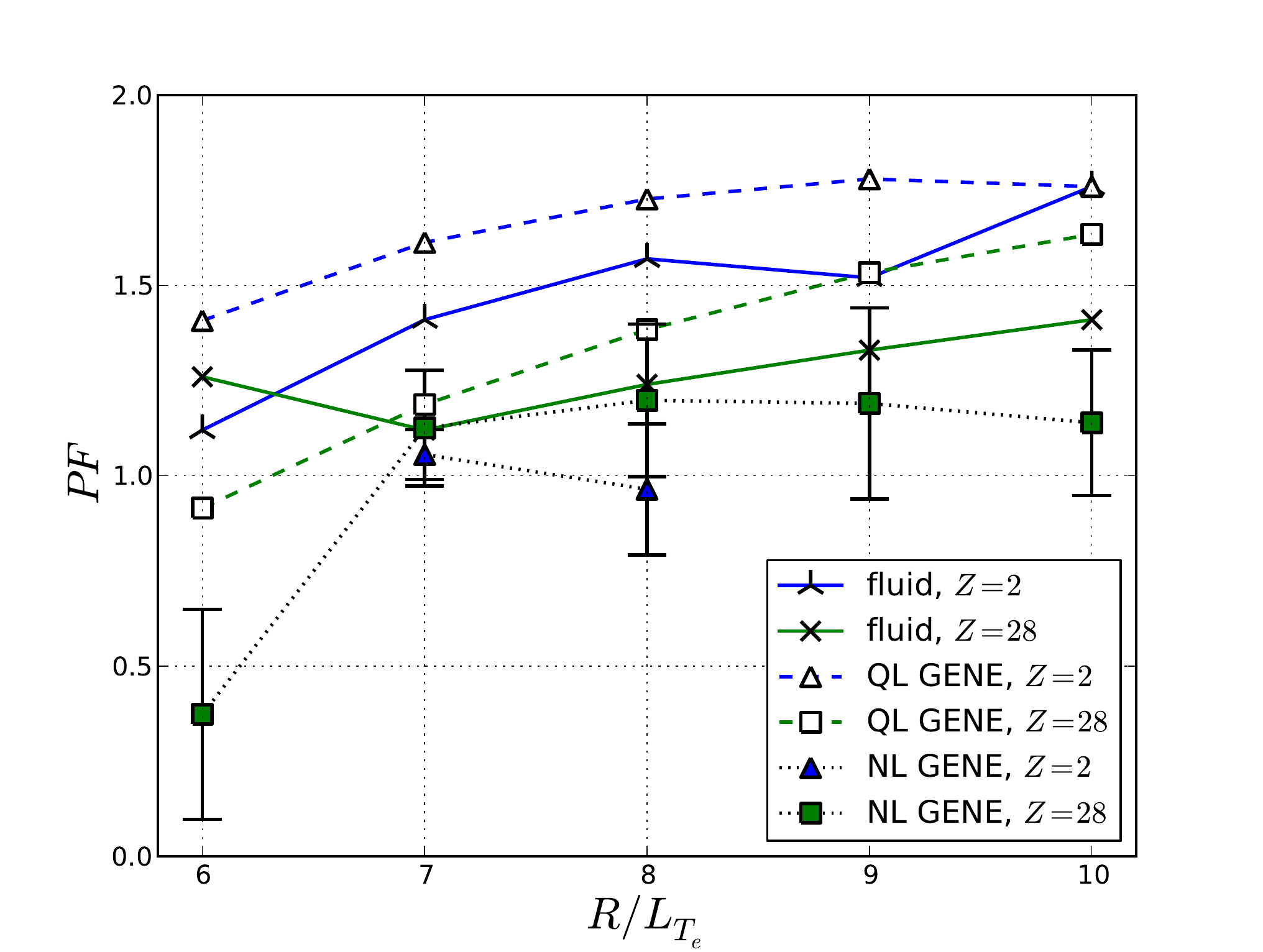}}
 
 \phantomcaption{}
\end{figure}

\begin{figure} %fig:omt_ITG
 \ContinuedFloat
 \centering
 \subfloat[dependence of the peaking factor ($PF$) on the normalised ion temperature gradient for the ITG~case; \label{fig:omt_ITG}]
 {\includegraphics[width=\figwidth]{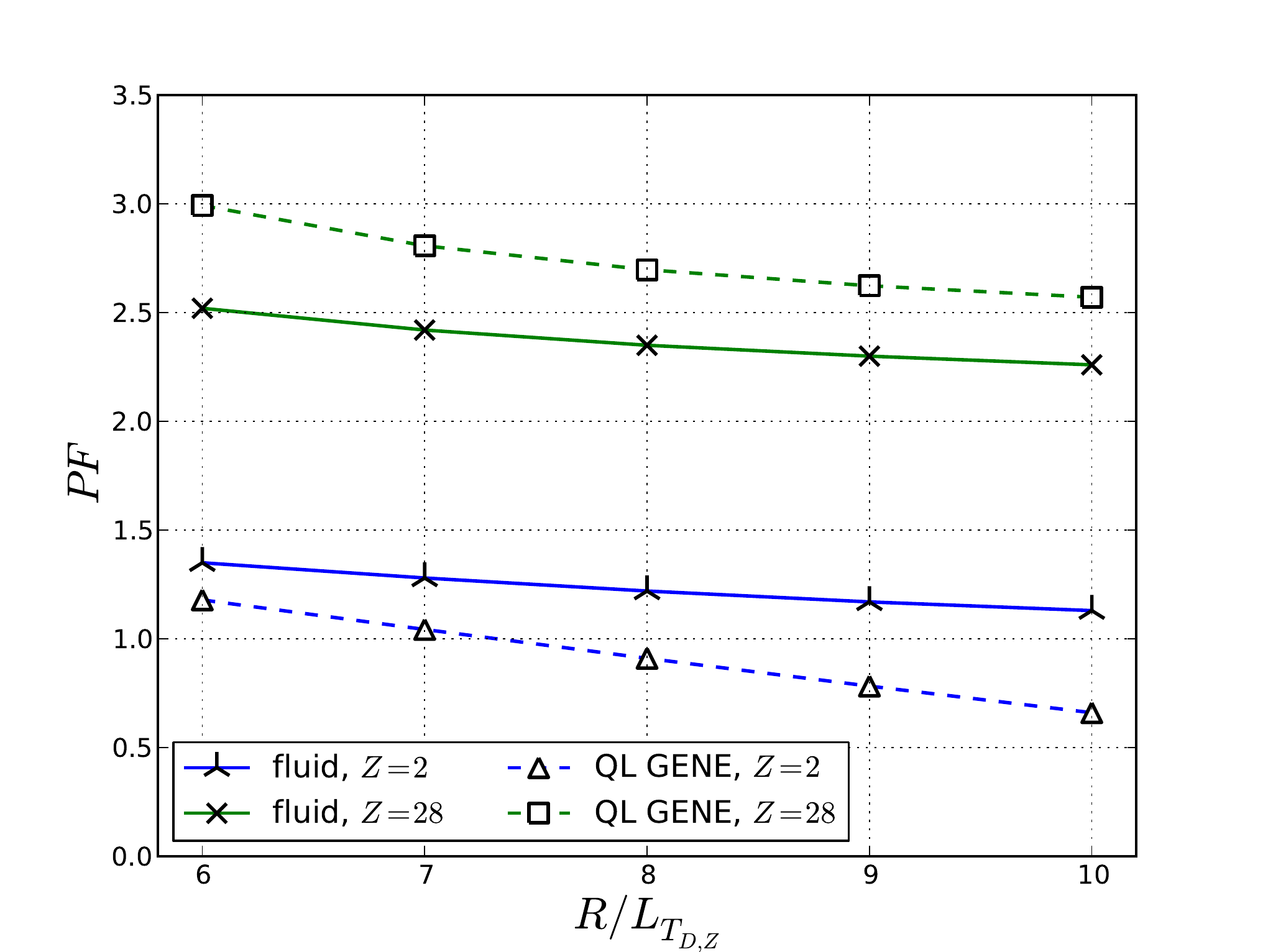}}
 
 \phantomcaption{}
\end{figure}

\begin{figure} %fig:omt_eigens, fig:omt
 \ContinuedFloat
 \centering
 \subfloat[real frequency ($\omega_r$) and growth rate ($\gamma$) for the two cases in Fig.~\ref{fig:omt_TEM} and Fig.~\ref{fig:omt_ITG} \label{fig:omt_eigens}]
 {\includegraphics[width=\figwidth]{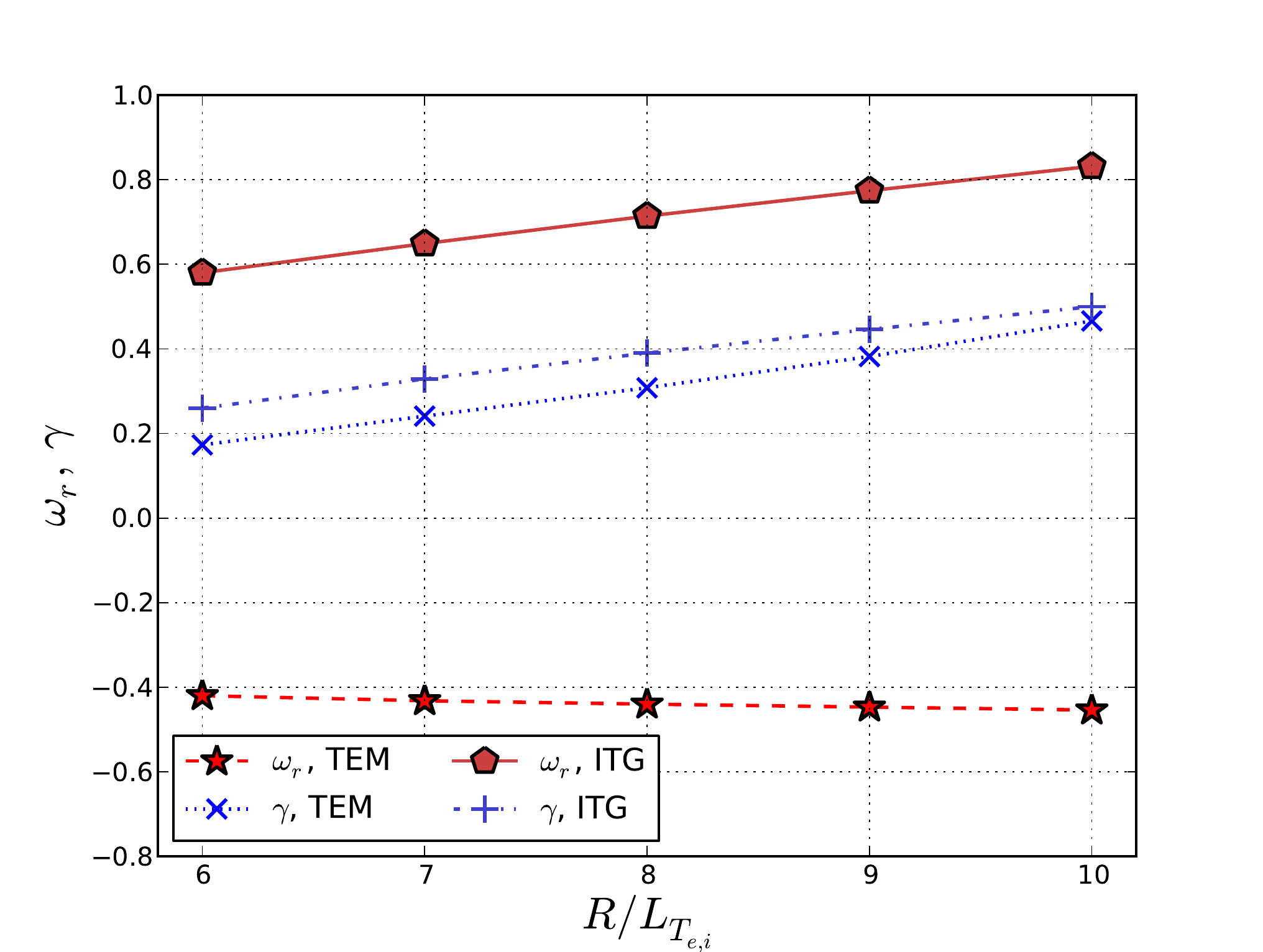}}
 
 \caption{Scalings of the peaking factor ($PF$) with the electron and ion temperature gradients ($-R\grad T_{e,i}/T_{e,i}=R/L_{T_{e,i}}$).
 Parameters for the TE~and ITG~mode case as in Fig.~\ref{fig:Z}, with $k_\theta\rho_s=0.2$.
 The eigenvalues in Fig.~\ref{fig:omt_eigens} are from QL~GENE simulations, they are normalised to $c_s/R$.}
 \label{fig:omt}
\end{figure}

\clearpage

\begin{figure}[\figplacing] %fig:omn_TEM
 \centering
 \subfloat[dependence of the peaking factor ($PF$) on the normalised electron density gradient for the TE~case, also indicated is the main ion peaking factor ($PF_e$) from fluid theory \label{fig:omn_TEM}]
 {\includegraphics[width=\figwidth]{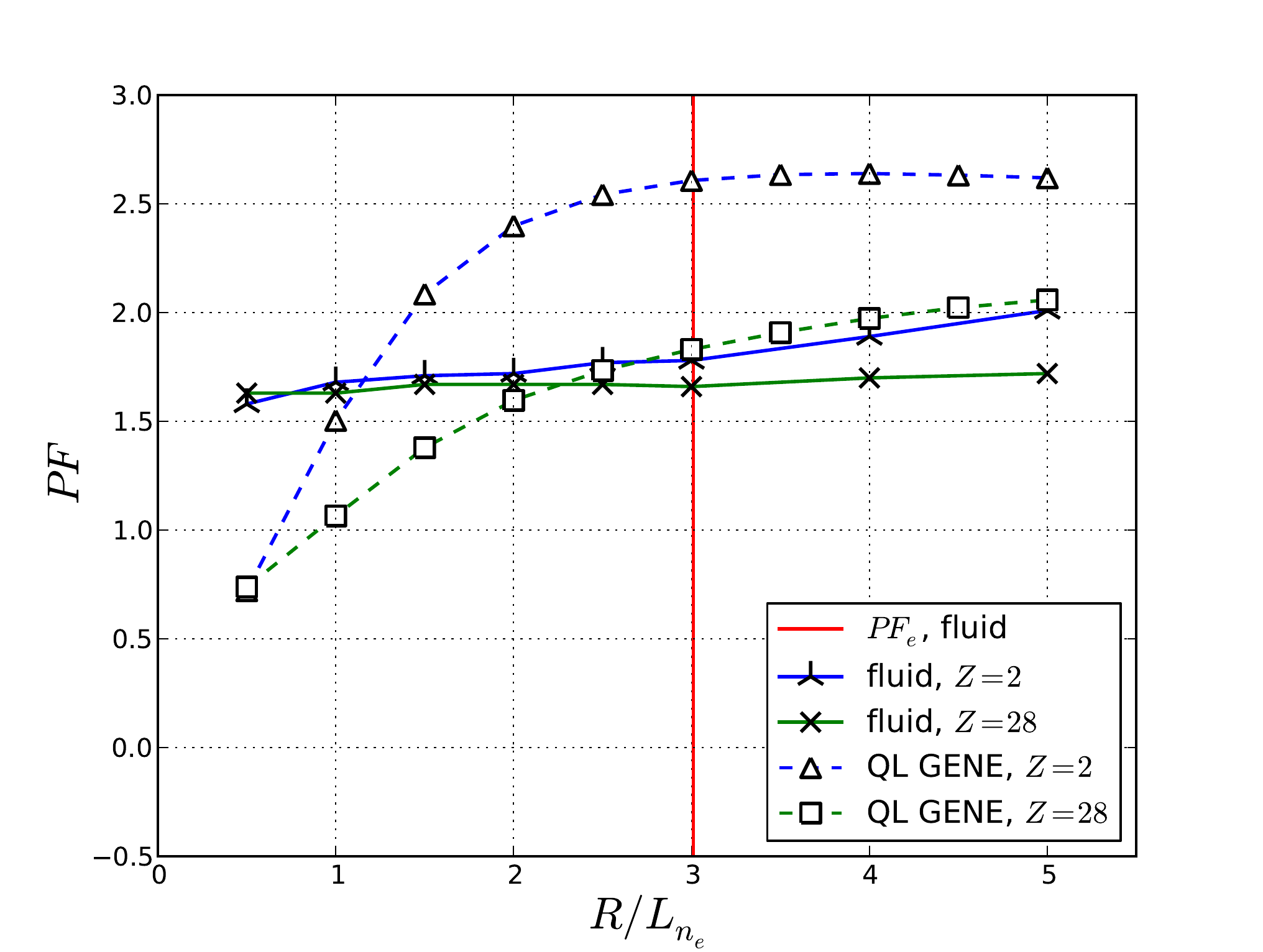}}
 
 \phantomcaption{}
\end{figure}

\begin{figure} %fig:omn_ITG
 \ContinuedFloat
 \centering
 \subfloat[dependence of the peaking factor ($PF$) on the normalised electron density gradient for the ITG~case, also indicated is the main ion peaking factor ($PF_e$) from fluid theory \label{fig:omn_ITG}]
 {\includegraphics[width=\figwidth]{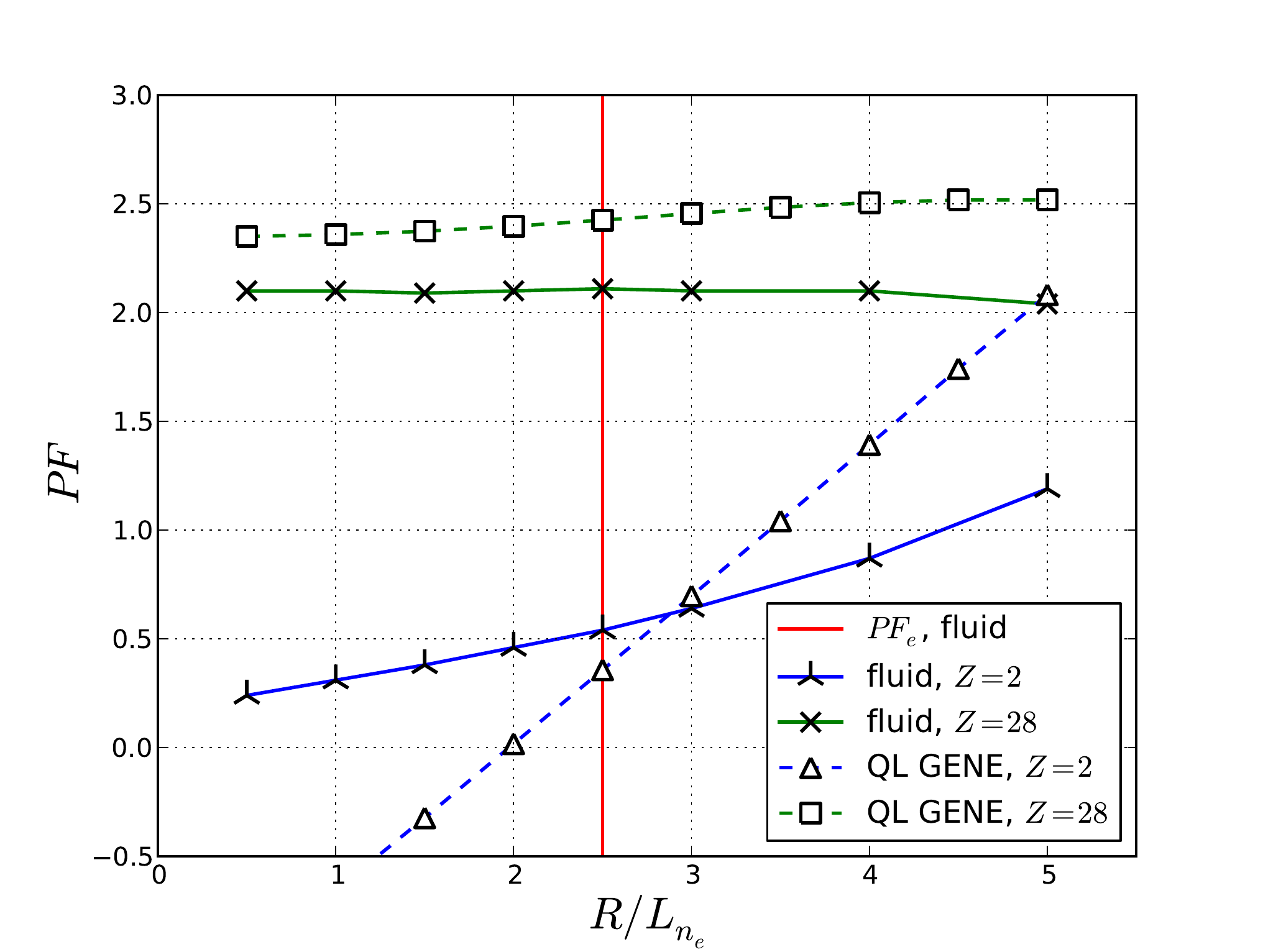}}
 \phantomcaption{}
\end{figure}

\begin{figure} %fig.omn_eigens, fig:omn
 \ContinuedFloat
 \centering
 \subfloat[real frequency ($\omega_r$) and growth rate ($\gamma$) for the two cases in Fig.~\ref{fig:omn_TEM} and Fig.~\ref{fig:omn_ITG} \label{fig:omn_eigens}]
 {\includegraphics[width=\figwidth]{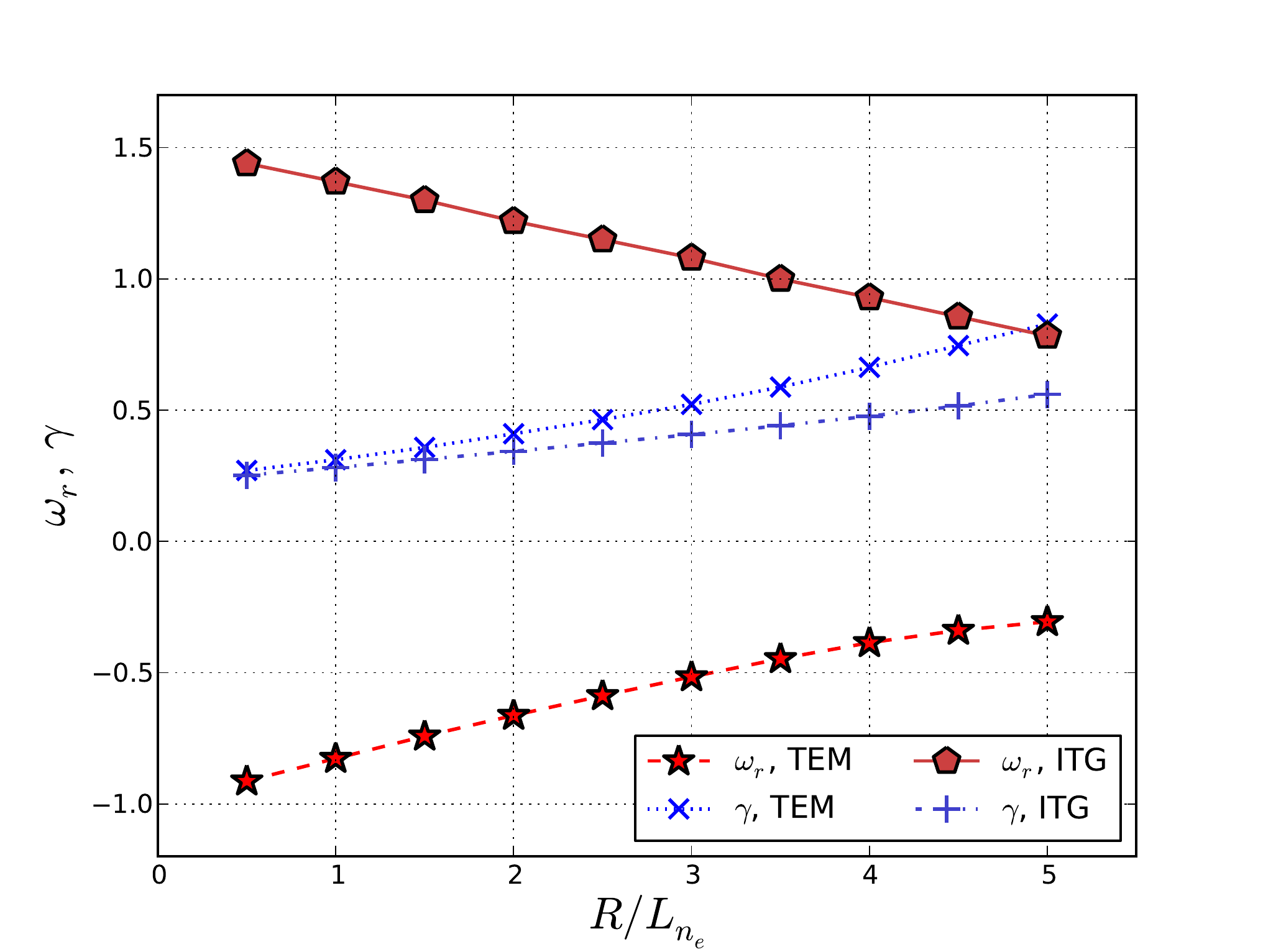}}
 
 \caption{Scalings of the peaking factor ($PF$) with the electron density gradient ($-R\grad n_e/n_e=R/L_{n_{e}}$).
 Parameters for the TE~and ITG~mode cases as in Fig.~\ref{fig:Z}, with $k_\theta\rho_s=0.3$ for both cases. %; H~plasma with trace impurities.
 The eigenvalues in Fig.~\ref{fig:omn_eigens} are from QL~GENE simulations, they are normalised to $c_s/R$.}
 \label{fig:omn}
\end{figure}

\end{document}